\documentclass{aa}  

\usepackage{tabularx}
\newcolumntype{Y}{>{\centering\arraybackslash}X} 
\usepackage{longtable}
\usepackage{supertabular}
\usepackage{graphicx}
\usepackage{siunitx}
\usepackage{caption}
\usepackage{subcaption} 
\usepackage{txfonts}
\usepackage[colorlinks=true,allcolors=blue]{hyperref}

\begin{document}

   \title{Tracing the sulfur depletion in starless and pre-stellar cores }

\author{L. Schöller\thanks{Corresponding author; \texttt{lausch@mpe.mpg.de}}\inst{1}\and S. Spezzano\inst{1} \and O. Sipilä\inst{1} \and E. I. Makarenko\inst{1} \and
P. Caselli\inst{1}\and H. A. Bunn\inst{1} \and S. S. Jensen\inst{1}}

\institute{Max-Planck-Institut f\"ur Extraterrestrische Physik, Giessenbachstrasse 1, 85748 Garching, Germany}

   \date{Received March 23, 2026; accepted May 6, 2026}

\abstract{Sulfur is one of the most abundant elements in the Universe, yet the sulfur budget inferred from the observed sulfur-bearing molecules in dense cores is significantly lower than expected. Starless and pre-stellar cores represent the earliest stages of star formation, thus providing an ideal laboratory to study the physical and chemical processes that cause sulfur depletion.}{We aim to constrain the sulfur chemistry in dense cores by measuring the abundances of different sulfur-bearing molecules and examining how these abundances reflect core evolution and environmental effects.}{We observed nine cores in the Taurus Molecular Cloud, targeting 13 sulfur-bearing molecules, including CS, CCS, C$_3$S, OCS, SO, SO$_2$, H$_2$CS, and various isotopologs. The molecular abundances and six abundance ratios were then compared to three evolutionary tracers: H$_2$ column density, N$_2$D$^+$/N$_2$H$^+$, and the CO depletion factor. We also compared the observed abundances with 0D chemical models with different initial sulfur abundances.}{We observe substantial variations in the abundances of individual sulfur-bearing molecules across the cores. L1517B exhibits consistently low abundances and a high depletion factor, whereas L1495B shows high abundances in oxygen-bearing species compared to the other cores within the L1495 filament. Ratios that probe the balance between carbon- and oxygen-bearing sulfur species (CCS/$^{34}$SO and C$^{34}$S/$^{34}$SO) decrease with increasing  H$_2$ column density and N$_2$D$^+$/N$_2$H$^+$ ratio. In contrast, individual molecules and other ratios show weak or no correlation with the evolutionary tracers. The 0D chemical models reproduce the abundances of some molecules, such as OCS, H$_2$CS, and HDCS, reasonably well but cannot simultaneously account for all observed species. This difference is most evident between the carbon- and oxygen-bearing molecules. }{The observed variations in sulfur abundances between the different cores and the lack of clear correlations with three standard evolutionary tracers indicate that a single evolutionary parameter cannot describe the sulfur chemistry. Instead, it is strongly influenced by local environmental factors. Reproducing the full sample of sulfur-bearing molecules would require improved chemical networks and models that account for the core's physical structure.}


   \maketitle

\section{Introduction}

Molecular clouds are known to be the birthplaces of stars. Their structure consists of a complex network of elongated filaments produced by the interplay of turbulence, magnetic fields, and self-gravity (e.g., \citealt{Andre2014}). The gravitational fragmentation of these filaments leads to the formation of dense substructures known as starless cores, which do not yet contain an embedded protostar (e.g., \citealt{Myers2009}). As their central densities increase ($n_\text{H} > 10^4$ cm$^{-3}$) and their temperatures remain low  (T < 10 K), a subset of these starless cores becomes gravitationally bound. Once the core becomes gravitationally unstable, it enters the pre-stellar phase before undergoing gravitational collapse (e.g., \citealt{Crapsi2005}).

At the pre-stellar stage, most of the material that will later form stars and planetary systems is already present. While this concept has been demonstrated for water through both observations and models, it has only been suggested by observations for complex organic molecules (\citealt{CaselliCeccarelli2012}; \citealt{Ceccarelli2014}; \citealt{Cleeves2014}; \citealt{Drozdovskaya2021}; \citealt{vanGelder2020}; \citealt{Jensen2021a}; \citealt{Jensen2021b}; \citealt{Scibelli2021}; \citealt{Scibelli2025}). Starless and pre-stellar cores therefore provide a unique window into the initial conditions just before the collapse and are essential for constraining the chemical budget inherited by forming stars and planetary systems (\citealt{Herbst2009}). 

In dense cores, the chemistry is dominated by an interplay between gas-phase reactions and grain-surface processes. The gas-phase chemistry is mainly driven by ion-molecule reactions, which are efficient at the low temperatures of dense cores. With the evolution of the core, the central density increases, leading to a decrease in temperature and the freezing-out of gas-phase molecules onto dust grains, which thus form icy mantles. Freeze-out onto dust grains was first observed for CO and CS (e.g., \citealt{Caselli1999}; \citealt{Tafalla2002}; \citealt{Tafalla2006Sulfurouterlayer}). Despite this depletion, the remaining gas-phase molecules still trace the ongoing chemistry. The resulting changes in molecular abundances make species such as CO and deuterated molecules powerful tracers of core evolution. CO is the main destruction partner of H$_3^+$. The freeze-out of CO promotes the reaction of H$_3^+$ with HD to form H$_2$D$^+$, initiating efficient deuterium fractionation. A common tracer of deuterium fractionation is the N$_2$D$^+$/N$_2$H$^+$ ratio (e.g., \citealt{Crapsi2005}; \citealt{Emprechtinger2009}).

Sulfur is one of the most abundant elements in the Universe, with a cosmic S/H abundance of  $\sim 1.6 \times 10^{-5}$ (\citealt{Asplund2021}). Sulfur-bearing molecules are widely observed in star-forming regions and are used to probe their physical conditions, such as the presence of shocks (e.g., \citealt{Dutrey2024}; \citealt{Esplugues2013}; \citealt{Hatchell1998}; \citealt{Bachiller1997}; \citealt{ZhangYang2023}). Despite all the efforts in observations and theory (e.g., \citealt{Fuente2023}; \citealt{Wakelam2004}), sulfur chemistry remains one of the least understood areas in astrochemistry. In particular, the total sulfur abundance observed in pre-stellar cores represents only a small fraction (< 1\%) of the cosmic sulfur abundance observed in diffuse clouds (e.g., \citealt{Tieftrunk1994}; \citealt{Woods2015}). While part of the sulfur reservoir could be locked up in ices or refractory grain material (e.g., \citealt{Ruffle1999}; \citealt{Martin-Domenech2016}; \citealt{LaasCaselli2019}), the solid-phase sulfur molecules identified so far account for only a few percent of the total sulfur budget, indicating that the main carrier of sulfur remains unknown; a part could be locked up in refractory sulfur allotropes (\citealt{Shingledecker2020}; \citealt{Cazaux2022}) and salts (\citealt{Poch2020}; \citealt{Slavicinska2025}) that are not detectable via rotational spectroscopy. 

Observations of sulfur-bearing molecules hint that sulfur depletion is linked to the evolutionary stage of the core. \citet{Nagy2019} find that the young starless core L1521E exhibits higher sulfur abundance levels than the more dynamically evolved pre-stellar core L1544. The difference in abundance depends on the sulfur-bearing molecules, with C$_3$S being roughly ten times more abundant in the more dynamically evolved core compared to the starless core. This trend suggests that a significant fraction of sulfur is depleted as cores evolve, consistent with the predictions by \citet{LaasCaselli2019}. \citet{Hily-Blant2022} confirmed this by using the observed NS/N$_2$H$^+$ ratio to infer the abundance of the gas-phase atomic sulfur from a sample of four starless and pre-stellar cores, and support a scenario in which sulfur becomes progressively depleted as dense cores evolve. 

Whether this behavior represents a general evolutionary trend or depends on the specific tracers used remains unclear. In this work, we observed various sulfur-bearing molecules in a sample of seven starless and two pre-stellar cores to investigate whether different evolutionary indicators trace sulfur depletion. Section \ref{sec:Observations} introduces the source sample, observational setup, instruments, and configurations used, and Sect. \ref{sec:Methods} outlines the derivation of column densities and the determination of the CO depletion factor. In Sect. \ref{sec:Results} we compare the observed abundances and selected molecular ratios according to different evolutionary tracers. We compare the observational abundances with those from chemical models, spanning different initial sulfur abundances, in Sect. \ref{sec:Models}. Finally, in Sects. \ref{sec:Discussion} and \ref{sec:Conclusion} we discuss the implications of our findings and conclude.

\section{Observations}
\label{sec:Observations}

Using the IRAM 30m telescope, situated on the Pico Veleta in Spain, we targeted the dust peaks of seven starless and two pre-stellar cores. All sources are located in the Taurus Molecular Cloud. Table \ref{tab:2_cores} summarizes the observed objects, including their coordinates.

\renewcommand{\arraystretch}{1.3}
\begin{table*}
    \centering
    \caption{\footnotesize Observed sources, classifications, corresponding coordinates, and CO depletion factors.}
    \vspace{-2mm} 
    \begin{tabularx}{\textwidth}{XXXcccc}
    \hline\hline
    Source & Class.$^\text{a}$ & R.A. (J2000) & Dec (J2000) & Ref. & $N$(H$_2$) (10$^{21}$ cm$^{-2}$)$^\text{b}$ & f$_\text{d}$ (C$^{17}$O)$^\text{c}$\\
    \hline
    CB23        & S  & 04:43:27.7 & 29:39:11.0 & 1 &  8.5 $\pm$ 1.6 & 1.05 $\pm$ 0.82\\
    L1495       & S  & 04:14:08.2 & 28:08:16.0 & 2 &  8.3 $\pm$ 1.6 & 1.28 $\pm$ 0.25\\
    L1495A-N    & S  & 04:18:31.8 & 28:27:30.0 & 1 & 15.1 $\pm$ 2.7 & 1.16 $\pm$ 0.21\\
    L1495A-S    & S  & 04:18:41.8 & 28:23:50.0 & 1 & 22.1 $\pm$ 4.1 & 0.57 $\pm$ 0.49\\
    L1495B      & S  & 04:18:05.1 & 28:22:22.0 & 1 & 10.1 $\pm$ 2.0 & 0.55 $\pm$ 0.14\\
    L1512       & S  & 05:04:09.7 & 32:43:09.0 & 1 &  8.6 $\pm$ 1.6 & 1.16 $\pm$ 0.22\\
    L1517B      & S  & 04:55:18.8 & 30:38:04.0 & 1 & 11.5 $\pm$ 2.3 & 2.47 $\pm$ 0.50\\
    TMC2        & P  & 04:32:48.7 & 24:25:12.0 & 1 & 18.9 $\pm$ 3.3 & 1.72 $\pm$ 0.30\\
    L1544       & P  & 05:04:17.2 & 25:10:43.0 & 3 & 27.9 $\pm$ 5.1 & 2.36 $\pm$ 1.12\\
    \hline
    \end{tabularx}

    \tablefoot{$^\text{a}$ S and P denote starless and pre-stellar cores. $^\text{b}$ Values taken from \citet{Chantzos2018}. $^\text{c}$ CO depletion factors were derived from C$^{17}$O observations in Sect. \ref{subsec:CO_depletion_factor}.

(1) \citet{Lee2001}, (2) \citet{Tafalla2002}, (3) \citet{Ward-Thompson1999}}

    \label{tab:2_cores}

\end{table*}
\renewcommand{\arraystretch}{1.0}

We collected the data using the Eight MIxer Receiver (EMIR) with the E090 band. The individual subbands were connected to the fast Fourier Transform Spectrometer (FTS), offering a total spectral coverage of \SI{7.2}{GHz} and a resolution of \SI{50}{kHz} (0.15 km/s at \SI{100}{GHz}). We employed frequency switching with a frequency throw of $\pm$ \SI{3.9}{MHz} and observed the core sample over a noncontinuous frequency range between approximately \SI{82}{GHz} and \SI{114.5}{GHz}.

During the observations (project IDs: 103-15, 120-22, 073-23, and 020-25), the weather conditions remained stable, with the optical depth ($\tau$) consistently being between 0.1 and 0.2. The pointing of the source was checked at least every two hours and corrected on nearby planets or pulsars, with an accuracy of 1" to 4". The resulting spectra reached typical root mean square (rms) noise levels of 2-25 mK per \SI{50}{kHz} channel, depending on the source and frequency. The spectral intensities are given in units of the antenna temperature, $T^*_\text{A}$. This translates to the main beam temperature $T_\text{MB}$ via $T_\text{MB} = F_\text{eff}/B_\text{eff} \cdot  T^*_\text{A}$.
Here, $F_\text{eff}$ and $B_\text{eff}$ represent the forward efficiency, which was assumed to be 95\% for all transitions, and the beam efficiency, which changed with the frequency of the telescope. The beam efficiency was determined through interpolation using the data provided on the IRAM 30m web page\footnote{\url{https://publicwiki.iram.es/Iram30mEfficiencies}}. For a few lines in Table \ref{tab:2_transitions}, the updated list of efficiencies was used, namely: SO (2,2 - 1,1) in L1495A-S and L1495B, SO (2,3 - 1,2) in TMC2, SO$_2$ (3$_{1,3}$ -  2$_{0,2}$) in TMC2, $^{13}$CS (2 - 1) in L1512 and L1517B,  CCS (7,6 - 6,5) in L1495A-S and L1495B, CCS (8,7 - 7,6) in CB23, H$_2$CS (3$_{1,3}$ - 2$_{1,2}$) in L1495A-S and L1495, and H$_2$CS (3$_{0,3}$ - 2$_{0,2}$) in TMC2. 

Within the observed frequency range, we focused on the sulfur-bearing molecules. All detected species contain hydrogen, carbon, and/or oxygen, in addition to sulfur. Table \ref{tab:2_transitions} summarizes the detected molecules and their spectroscopic information: $E_\text{u}$ represents the upper level energy, $Q_\text{rot}$ the rotational partition function, $g_\text{u}$ the upper level degeneracy, and $A_\text{ul}$ the Einstein coefficient for spontaneous emission.

\renewcommand{\arraystretch}{1.3}
\begin{table*}[]
    \centering
    \caption{Transition parameters of the observed molecules and setting of the telescope.}
    \vspace{-2mm} 
    \begin{tabularx}{\textwidth}{lXcccccccc}
        \hline\hline
        Molecule & Transition & Frequency & Reference &  $E_\text{u}$ &  $g_\text{u}$ & $A_\text{ul}$ & $Q_\text{rot}$ ($T_\text{ex}$)     & $n_\text{crit}$ ($T_\text{ex}$) & $F_\text{eff}$/$B_\text{eff}$$^\text{a}$ \\ 
                 &            & (MHz)     &           &        (K)     &            &(10$^{-5}$ s$^{-1}$) &  (K) & (10$^5$ cm$^{-3}$) & (\%) /(\%)\\
        \hline
        SO          & $N,J$ = 2,2 - 1,1               & 86093.95  & 1  & 19.31  & 5  &  0.525 & 17.28 & 1.57 & 95 /80.9 (81.5)\\
                    & $N,J$ = 2,3 - 1,2               & 99299.87  & 1  & 9.22   & 7  &  1.125 & 17.28 & 1.39 &95 /79.6 (76.9)\\
                    & $N,J$ = 3,2 - 2,1               & 109252.22 & 1  & 21.05  & 5  &  1.080 & 17.28 & 1.37 &95 /78.5\\
        S$^{18}$O   & $N,J$ = 2,3 - 1,2               & 93267.27  & 2  & 8.72   & 7  &  0.934 & 18.37 & 1.39 &95 /80.2\\
        $^{34}$SO   & $N,J$ = 2,3 - 1,2               & 97715.32  & 2  & 9.09   & 7  &  1.073 & 17.55 &1.39 & 95 /79.7\\
        SO$_2$      & $J_{\text{K}_\text{a},\text{K}_\text{c}}$ = 3$_{1,3}$ -  2$_{0,2}$  & 104029.42 & 3  & 7.74   & 7  &  1.006 & 36.39 & 2.14 &95 /79.1 (74.8)\\
        OCS         & $J$ = 8 - 7                   & 97301.21  & 4  & 21.02  & 17 &  0.258 & 34.60 & 0.35 & 95 /79.8\\
                    & $J$ = 9 - 8                   & 109463.06 & 4  & 26.26  & 19 &  0.370 & 34.60 & 0.49 &95 /79.1\\
        CS          & $J$ = 2 - 1               & 97980.95  & 5  & 7.05   & 5  &  1.679 &  8.85 & 3.58 &95 /79.7\\
        $^{13}$CS   & $J$ = 2 - 1               & 92494.31  & 5  & 6.65   & 5  &  1.413 &  9.35 & 3.58 &95 /80.2 (79.6)\\
        C$^{34}$S   & $J$ = 2 - 1     & 96412.95  & 5  & 6.93  & 5  &  1.600 &  8.98 & 3.58 &95 /79.8\\
        CCS         & $N,J$ = 7,6 - 6,5               & 86181.39  & 6  & 23.35  & 13 &  2.778 & 61.61 & 3.64 &95 /80.8 (81.4)\\
                    & $N,J$ = 7,8 - 6,7               & 93870.11  & 6  & 19.90  & 17 &  3.744 & 61.61 & 4.21 &95 /80.1\\
                    & $N,J$ = 8,7 - 7,6               & 99866.52  & 6  & 28.14  & 15 &  4.404 & 61.61 & 5.50 &95 /79.5 (76.4) \\
                    & $N,J$ = 8,8 - 7,7               & 103640.76 & 6  & 31.22  & 15 &  4.979 & 61.61 & 6.11 &95 /79.1\\
                    & $N,J$ = 9,8 - 8,7               & 113410.18 & 6  & 33.58  & 17 &  6.534 & 61.61 & 7.84 &95 /78.2\\
        C$_3$S      & $J$ = 16 - 15                 & 92488.49  & 7  & 37.72  & 33 &  6.126 & 72.42 & - &95 /80.2\\
        HCS$^+$     & $J$ = 2 - 1                   & 85347.89  & 8  & 6.14   & 5  &  1.110 & 10.11 & 0.29 &95 /81.0\\
        H$_2$CS     & $J_{\text{K}_\text{a},\text{K}_\text{c}}$ = 3$_{1,3}$ -  2$_{1,2}$  & 101477.80 & 9  & 22.99  & 21 &  1.260 & 31.01 & 1.59 &95 /79.4 (75.6)\\
                    & $J_{\text{K}_\text{a},\text{K}_\text{c}}$ = 3$_{0,3}$ -  2$_{0,2}$  & 103040.45 & 9  & 9.88   & 7  &  1.484 & 31.01 & 1.30 &95 /79.2 (75.1)\\
                    & $J_{\text{K}_\text{a},\text{K}_\text{c}}$ = 3$_{1,2}$ -  2$_{1,1}$  & 104617.03 & 9  & 23.21  & 21 &  1.381 & 31.01 & 2.26 &95 /79.0\\
        HDCS        & $J_{\text{K}_\text{a},\text{K}_\text{c}}$ = 3$_{0,3}$ -  2$_{0,2}$  & 92981.60  & 10 &8.92   & 7  &  1.097 & 24.77 & 1.30 &95 /80.2\\
        C$^{17}$O   & $J$ = 1 - 0                   & 112360.01  & 11 &   5.39    &   3   &   0.0067  & 4.06 & 0.02 &95/78.3    \\
                    \hline
            
    \end{tabularx}

\tablefoot{$E_\text{u}$: energy of the upper level; $g_\text{u}$: degeneracy of the upper level; $A_\text{ul}$: Einstein coefficient for spontaneous emission; $Q_\text{rot}$: rotational partition function at excitation temperature $T_\text{ex} = 10$; $n_\text{crit}$: critical density at excitation temperature $T_\text{ex} = 10$. All values were taken or calculated from the CDMS \citep{CDMS} or JPL \citep{Pickett1998JPL} catalogs, except the collisional rates, which were taken from LAMBDA \citep{LAMBDA} or EMAA \citep{EMAA}. Note that collisional rate coefficients for all isotopologs are taken to be equal to those of the main isotopic species, as no corresponding data are available in the databases.
$^\text{a}$ Values in brackets represent the new beam efficiencies as of Dec. 2025 where needed.
    
\textbf{References.} (1) \citet{transitionsSO}; (2) \citet{transitionsSOiso}; (3) \citet{transitionsSO2}; (4) \citet{transitionsOCS}; (5) \citet{transitionsCS}; (6) \citet{transitionsCCS}; (7) \citet{transitionsC3S}; (8) \citet{transitionsHCSpl}; (9) \citet{transitionsH2CS}; (10) \citet{transitionsHDCS}; (11) \citet{transitionsC17O}}

    \label{tab:2_transitions}
\end{table*}
\renewcommand{\arraystretch}{1.0}

\section{Derivation of the column density and CO depletion factor }
\label{sec:Methods}

\subsection{Molecular column densities}
The initial reduction of the observational data were performed using the \texttt{GILDAS} software (\citealt{Pety2005Gildas}). To correct for the baseline caused by frequency switching, we fitted and subtracted a third-degree polynomial. The individual lines were then fitted with a Gaussian line profile using \texttt{python}. Figure \ref{fig:3_linesL1495AS} shows the spectra of L1495A-S with the Gaussian fits overlaid. The spectra of the remaining cores are shown in Appendix \ref{sec:AppendixA}. For the calculation of the total column density $N_\text{tot}^\text{thin}$, we employed the formula reported in \citet{MangumShirleycoldens} for the optically thin emission under the assumption of local thermal equilibrium (LTE) and a constant excitation temperature ($T_\text{ex}$) of \SI{10}{K}:

\begin{figure*}[h!]
    \centering
    \includegraphics[width=\textwidth]{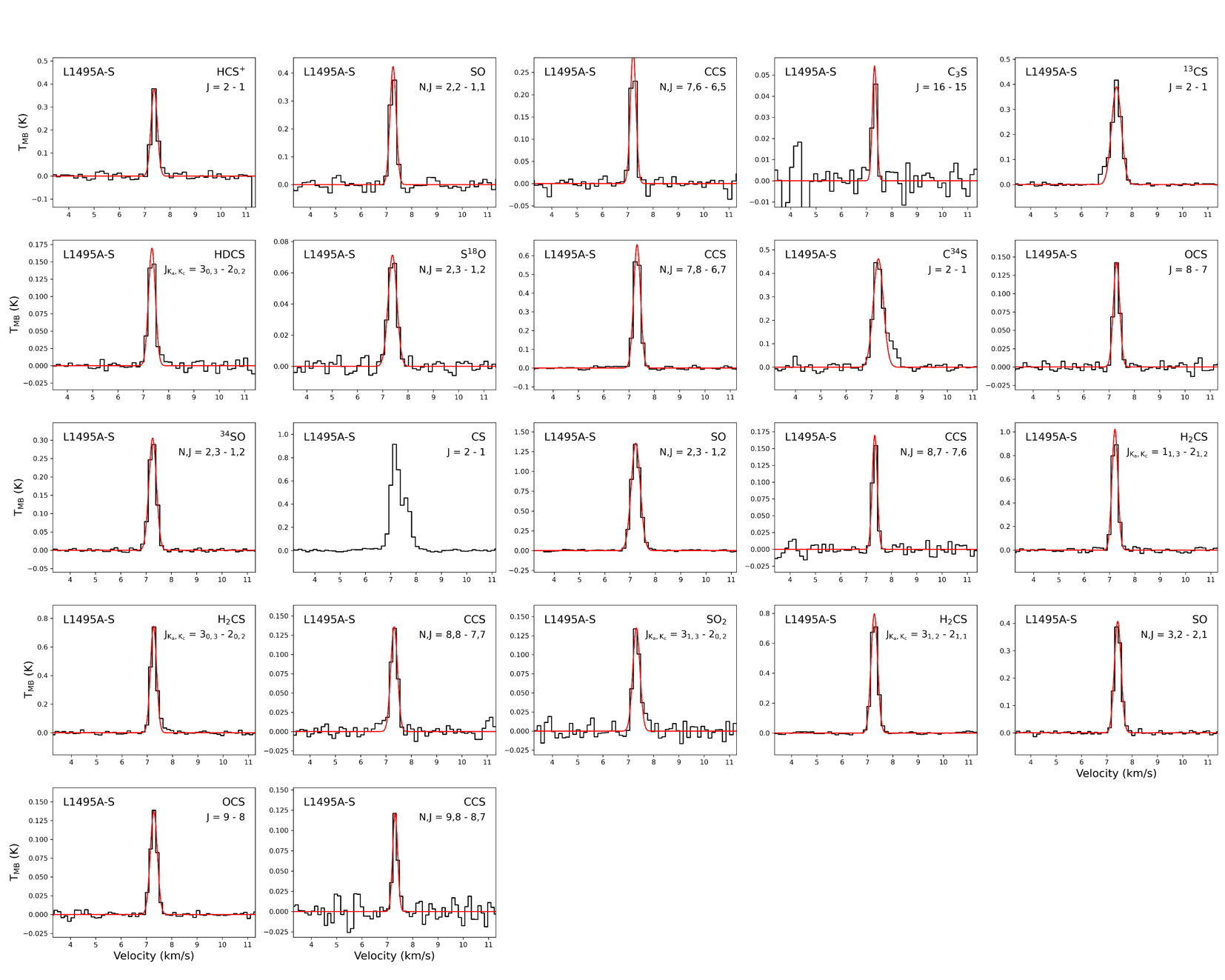}
    \caption{Spectra of molecules observed toward the dust peak of the starless core L1495A-S (in black). The red line indicates the best fit to a Gaussian line shape model. For each panel, the molecule is indicated on the top right, with the transition stated below. CS shows signs of self-absorption; hence, no Gaussian fit was performed. }
    \label{fig:3_linesL1495AS}
\end{figure*}

\begin{equation}
    N_\text{tot}^\text{thin} = \frac{8 \pi k_\text{B} \nu^2 }{hc^3 } \frac{Q_\text{rot}(T_\text{ex})}{g_\text{u}A_\text{ul}} \frac{J(T_\text{ex})}{J(T_\text{ex})-J(T_\text{bg})} \text{   }e^{E_\text{u}/k_\text{B}T_\text{ex}}  \int T_\text{mb} \text{dv},
    \label{eq:3_coldens}
\end{equation}

\noindent
where $\nu$ represents the transition frequency, $ k_\text{B}$ is the Boltzmann constant, $h$ is the Planck constant and $c$ is the speed of light. $J(T) = \frac{h\nu}{k_\text{B}} (e^{h\nu/k_\text{B}T}-1)^{-1}$ is the Rayleigh-Jeans equivalent temperature and $T_\text{bg}$= \SI{2.73}{K} is the temperature of the cosmic microwave background. When assuming a Gaussian line profile, the integrated main beam temperature was determined via $\frac{1}{2} \sqrt{\pi/ln(2)} \Delta$v $\cdot T_\text{mb}$, with $\Delta$v describing the full width half maximum of the line. To account for any deviations of the optically thin regime, we determined the optical depth ($\tau$) and scaled the column density with the optical depth correction factor (\citealt{GoldsmithLangercorrectfactor}) via
\begin{equation}
    N_\text{tot} = N_\text{tot}^\text{thin} \frac{\tau}{1-e^{-\tau}},
    \label{eq:3_correctionfactor}
\end{equation}

\noindent with
\begin{equation}
    \tau = - \ln \left[ 1- \frac{T_\text{mb}}{f [J(T_\text{ex}) - J (T_\text{bg}) ]} \right].
\end{equation}

\noindent Here, $f$ describes the beam filling factor, which we set to 1 for all cores. For the excitation temperature, we assumed \SI{10}{K} for all molecules. Although the innermost region of a core can be a few kelvin lower, the envelopes are well characterized by temperatures of around \SI{10}{K}, where most of the sulfur-bearing species are located (e.g., \citealt{Tafalla2006Sulfurouterlayer}). Varying the excitation temperature by $\pm$ \SI{1}{K} had a negligible effect on the derived column density for all molecules except SO, which showed a change of up to 10\%. Given that the critical densities of the observed transitions (see Table \ref{tab:2_transitions}) are lower or comparable to typical densities of the core, LTE provides a reasonable first-order approximation for describing the observed emission. While this may not hold for all transitions in all cores, the limited number of detected optically thin transitions per molecule does not provide enough constraints for a meaningful non-LTE analysis. We therefore used an approach with a uniform excitation temperature and LTE to ensure a homogeneous comparison across all species. 

Table \ref{tab:3_transitons_L1495AS} reports the Gaussian fit parameters, the optical depth, and the corrected column densities for all the detected lines for L1495A-S. The information of the remaining cores can be found in Appendix \ref{sec:AppendixB}. For non-detections, we determined upper limits for the integrated intensity, which were further used to derive upper limits on the column density. The uncertainties for the integrated intensities and linewidths were taken from the Gaussian fit. For the column densities under the optically thin assumption, the uncertainties were determined by propagating the uncertainty of the integrated intensity. For the uncertainty of $\tau$ and the corrected total column density, we then used Gaussian error propagation.

\renewcommand{\arraystretch}{1.3}

\begin{table*}[]
    \caption{Line properties of L1495A-S, showing Gaussian fit results for the observed transitions and upper limits for non-detections.}
    \vspace{-3mm}
    \centering
    \begin{tabularx}{\textwidth}{lXXccXXcc}
        \hline\hline
        Source / & Frequency & T$_\text{MB}$ & rms &  W  &  v$_\text{LSR}$ &$\Delta$v & $\tau$ & N$_\text{tot}$ \\ 
        Molecule      &  (GHz)    & (K) & (mK)&(K km s$^{-1}$)& (km s$^{-1}$) &(km s$^{-1}$) &  &($10^{11}$ cm$^{-2}$) \\
        \hline        
         \large \textbf{\textit{L1495A-S}} &  &  &  &  &  & & & \\
        HCS$^+$     & 85.347 & 0.38       &  10  & 0.118 $\pm$   0.013 & 7.375  & 0.293  &        0.056 $\pm$     0.006             & 6.8 $\pm$   0.7  \\
        SO          & 86.093 & 0.43       &  14  & 0.130 $\pm$   0.005 & 7.267  & 0.288  &        0.063 $\pm$     0.002             & 102.6 $\pm$         3.6   \\
        CCS         & 86.181 & 0.30       &  11  & 0.083 $\pm$   0.004 & 7.217  & 0.261  &        0.044 $\pm$     0.003             & 25.2 $\pm$  1.2 \\
        C$_3$S      & 92.488 & 0.05       &  6   & 0.012 $\pm$   0.002 & 7.314  & 0.203  &        0.008 $\pm$     0.002             & 3.5 $\pm$   0.5\\
        $^{13}$CS   & 92.494 & 0.39       &  9   & 0.190 $\pm$   0.004 & 7.389  & 0.457  &        0.065 $\pm$     0.002             & 9.7 $\pm$   0.2 \\
        HDCS        & 92.981 & 0.17       &  8   & 0.058 $\pm$   0.003 & 7.344  & 0.319  &        0.025 $\pm$     0.001             & 9.0 $\pm$   0.4 \\
        S$^{18}$O   & 93.267 & 0.07       &  3   & 0.031 $\pm$   0.001 & 7.385  & 0.406  &        0.011 $\pm$     0.000             & 4.1 $\pm$   0.2 \\
        CCS         & 93.870 & 0.66       &  5   & 0.223 $\pm$   0.002 & 7.317  & 0.318  &        0.100 $\pm$     0.001             & 32.8 $\pm$  0.3\\       
        C$^{34}$S   & 96.412 & 0.46       &  14  & 0.232 $\pm$   0.007 & 7.273  & 0.471  &        0.078 $\pm$     0.003             & 11.3 $\pm$  0.4 \\
        OCS         & 97.301 & 0.14       &  6   & 0.045 $\pm$   0.002 & 7.298  & 0.296  &        0.021 $\pm$     0.001             & 62.1 $\pm$  2.2 \\
        $^{34}$SO   & 97.715 & 0.31       &  4   & 0.115 $\pm$   0.001 & 7.264  & 0.352  &        0.045 $\pm$     0.001             & 14.5 $\pm$  0.2 \\
        SO          & 99.299 & 1.36       &  8   & 0.610 $\pm$   0.003 & 7.223  & 0.422  &        0.223 $\pm$     0.015             & 82.7 $\pm$  0.7 \\
        CCS         & 99.866 & 0.17       &  7   & 0.042 $\pm$   0.002 & 7.345  & 0.233  &        0.025 $\pm$     0.001             & 14.8 $\pm$  0.6\\
        H$_2$CS     &101.477 & 1.024      &  15  & 0.300 $\pm$   0.005 & 7.215  & 0.275  &        0.164 $\pm$     0.003             & 87.1 $\pm$  1.4\\
        H$_2$CS     &103.040 & 0.75       &  10  & 0.230 $\pm$   0.004 & 7.247  & 0.290  &        0.116 $\pm$     0.002             & 46.1 $\pm$  0.7 \\
        CCS         &103.640 & 0.14       &  5   & 0.046 $\pm$   0.002 & 7.321  & 0.315  &        0.020 $\pm$     0.001             & 20.6 $\pm$  0.9 \\
        SO$_2$      &104.029 & 0.14       &  10  & 0.048 $\pm$   0.003 & 7.321  & 0.334  &        0.020 $\pm$     0.001             & 13.1 $\pm$  0.8 \\
        H$_2$CS     &104.617 & 0.80       &  7   & 0.262 $\pm$   0.002 & 7.272  & 0.309  &        0.124 $\pm$     0.001             & 73.6 $\pm$  0.6 \\
        SO          &109.252 &  0.41      &  6   & 0.137 $\pm$   0.002 & 7.431  & 0.315  &        0.062 $\pm$     0.001             & 97.5 $\pm$  1.3 \\
        OCS         &109.463 & 0.14       &  5   & 0.048 $\pm$   0.001 & 7.294  & 0.326  &        0.020 $\pm$     0.001             & 86.7 $\pm$  2.5\\
        CCS         &113.410 & 0.12       &  14  & 0.030 $\pm$   0.003 & 7.312  & 0.232  &        0.018 $\pm$     0.002             & 13.7 $\pm$  1.4 \\
        \hline
    \end{tabularx}
    \label{tab:3_transitons_L1495AS}
    
\tablefoot{Line properties were derived using Gaussian fits. Upper limits were determined as $3 \sigma \sqrt{\Delta \text{v} \delta\text{v}}$, following \citet{JimenezSerraupperlimits2016}, where $\sigma$ is the rms noise, $\Delta$v the full width at half maximum, and $\delta$v the spectral velocity resolution. The reported v$_\text{LSR}$ and $\Delta$v values are given without uncertainties, as the statistical uncertainties are significantly smaller than the spectral resolution.}

\end{table*}
\renewcommand{\arraystretch}{1.0}

CS, one of the most abundant sulfur-bearing species in starless and pre-stellar cores, shows clear signs of high optical depth (e.g., self-absorption) in L1495A-S, L1495A-N, and L1544 (Figs. \ref{fig:3_linesL1495AS}, \ref{fig:A_linesL1495AN}, and \ref{fig:A_linesL1544}) and was therefore excluded from further analysis in these sources. In L1495B (Figure \ref{fig:A_linesL1495B}), wing-like structures affect the CS (2-1) and SO (2,3-1,2) transitions, leading to their exclusion. The weak transitions owing to OCS (8-7), SO$_2$ (3$_{1,3}$ -  2$_{0,2}$) in TMC2 (Figure \ref{fig:A_linesTMC2}), HDCS (3$_{0,3}$ -  2$_{0,2}$) in L1495 (Figure \ref{fig:A_linesL1495}), OCS (8-7) in L1495A-N (Figure \ref{fig:A_linesL1495AN}), and HCS$^+$ (2-1) in L1517B (Figure \ref{fig:A_linesL1517B}) show features resembling a second velocity component over one to two channels. Given their low S/R, these features were attributed to noise and do not influence the Gaussian fits. Therefore, only HCS$^+$ in L1517B was excluded due to a linewidth, which is $\sim$40\% broader than the other lines in that core. Additionally, four molecular transitions in each TMC2 and L1495A-S (Figs. \ref{fig:A_linesTMC2} and \ref{fig:3_linesL1495AS}) show one-sided wings, which may indicate infall motions (e.g., \citealt{Lee1999}; \citealt{Lee2001}; \citealt{Crapsi2005}). The presence of both red- and blueshifted asymmetries, which vary between molecular tracers, suggests that the features could arise from different regions within the cores or reflect different evolutionary stages. To further constrain their origin, maps to understand the spatial distribution and spectra with higher frequency resolution would be necessary. As the features were not accounted for in the Gaussian fit, the line parameters from the fit are unaffected and were thus included in the further analysis. The transitions with an asterisk in Table \hyperref[sec:AppendixB]{B.1} mark the excluded lines. 

Excluding the specified transitions, the average linewidths across the cores range from approximately \SI{0.3}{km/s} to \SI{0.45}{km/s}. The pre-stellar cores L1544 and TMC2 have the highest average linewidths, and L1512 and CB23 the narrowest. For most sources, the average linewidths agree within uncertainties; only the linewidths of L1544 and TMC2 do not agree with those of CB23, L1512, L1495A-S, and L1495B.

\subsection{CO depletion factor}
\label{subsec:CO_depletion_factor}
Throughout most of their evolution, dense cores maintain a temperature of around \SI{10}{K}. When their central number densities exceed $\sim 10^5$ cm$^{-3}$, molecular freeze-out becomes catastrophic, as gas-phase molecules collide with and stick to dust grains, drastically reducing their gas-phase abundances. This process, which reflects the evolutionary stage of the core, was first observed for CO (e.g., \citealt{Caselli1999}; \citealt{Tafalla2002}; \citealt{Willacy1998}). The degree of the CO freeze-out can be measured using the CO depletion factor $f_\text{d}$. This measure compares the observed CO abundance $X$(CO) with the CO abundance in the local ISM via (e.g., \citealt{Emprechtinger2009})\begin{equation}
    f_\text{d}(\text{CO}) = \frac{X_\text{ref} (\text{CO})}{X(\text{CO})}.
\end{equation}
For the reference CO abundance  $X_\text{ref} (\text{CO}),$ we used $9.5 \cdot 10^{-5}$ (\citealt{Frerking1982}). We derived the observed CO abundance from the C$^{17}$O ($J$ = 1-0) transition by comparing its total column density (see the derivation in Appendix \ref{sec:AppendixD}) with the H$_2$ column density of the core while assuming $X(\text{C}^{16}\text{O})$/ $X(\text{C}^{17}\text{O})$ = 2044 (\citealt{WilsonWood1994,Penzias1981}). To determine the observed CO abundance, we then used 
\begin{equation}
    X(C^{16}\text{O}) = \frac{\text{X}(\text{C}^{17}\text{O})\cdot 2044}{N(\text{H}_2)}.
\end{equation}
For all the starless and pre-stellar cores in the sample, we adopted the H$_2$ column densities derived within a 40" beam from \textit{Herschel}/SPIRE observations as reported in \citet{Chantzos2018}. The resulting derived CO depletion factors are shown in Table \ref{tab:2_cores}.

\section{Results}
\label{sec:Results}

To compare the abundances, we selected one transition per molecule across the core sample to ensure a consistent, nonredundant comparison. Whenever more than one transition was detected, for example CCS, the transition detected in most cores was selected. When multiple transitions had the same number of detections, for example, H$_2$CS, the transition with the higher observed intensity was taken. For SO, the N, J = 2,3 - 1,2 transition was used for all cores except L1495B due to optical depth effects. Because of high optical depth, CS was also excluded for four cores. Only five CS abundances remained, which limits the statistical significance of the analysis. Given that the isotopic species $^{13}$CS and C$^{34}$S, which we expect to trace similar regions as CS, are detected in all cores, we excluded the main isotopolog CS from the analysis. The transitions analyzed in this work are listed in Table \ref{tab:4_abundances}. 

The following section is divided into two parts: molecular abundances and selected abundance ratios used to probe chemical trends among sulfur-bearing species. In both parts, abundances and various molecular ratios are shown as a function of three evolutionary tracers: the H$_2$ column density (taken from \citealt{Chantzos2018}), the N$_2$D$^+$/N$_2$H$^+$ ratio (taken from \citealt{Crapsi2005}) and the CO depletion factor (derived in Sect. \ref{subsec:CO_depletion_factor}), in order to search for possible correlations. Potential trends were quantified with the Spearman rank correlation coefficient (including both detections and non-detections). For molecules or ratios showing a trend ($ |\rho| \geq$ 0.6 and $p \leq$ 0.1), a leave-one-out analysis was performed to identify possible outliers. Note that the small sample size of cores limits the statistical significance of the results, and the Spearman rank coefficient does not account for uncertainties in the observed abundances. This analysis was also applied to the correlations between the evolutionary tracers themselves. No strong correlations are found within the sample, with the strongest trend observed between the H$_2$ column density and the N$_2$D$^+$/N$_2$H$^+$ ratio, with a Spearman rank coefficient of $\rho =$0.57 and $p=$0.11.

\subsection{Abundances}
\label{subsec:4_Abundances}

The observational (fractional) abundances are defined as the ratio of the molecular column density and the column density of H$_2$. Table \ref{tab:4_abundances} shows the derived abundances for each molecule in the individual cores, with upper limits indicating non-detections. The highest abundances for all observed molecules are found in L1495A-S and L1495A-N, with L1495A-S showing the largest number of detected molecules together with L1544. L1517B exhibits the lowest abundances, and together with L1495, CB23, and L1512, displays the highest rate of non-detections. For these four cores, only 7 out of 12 molecules were detected.

\renewcommand{\arraystretch}{1.3}
\begin{table*}[ht]
    \centering
    \caption{Observed molecular abundances in the cores for all observed sulfur-bearing molecules.}
    \resizebox{\textwidth}{!}{%
    \begin{tabular}{cccccccccc}
    \hline\hline
       Molecular Abundances&   CB23    &  TMC2 & L1495 & L1495A-N  & L1495A-S  & L1495B    & L1512 & L1517B    & L1544\\
       &   $10^{-11}$  &   $10^{-11}$ &  $10^{-11}$ &  $10^{-11}$  &  $10^{-11}$  &  $10^{-11}$    &  $10^{-11}$ &  $10^{-11}$    &  $10^{-11}$\\
    \hline
    SO$^\text{a}$ & 8.58 $\pm$ 1.51 & 77.26 $\pm$ 14.44 & 11.18 $\pm$ 2.14 & 56.01 $\pm$ 11.11 & 71.88 $\pm$ 14.39 & 60.04 $\pm$ 10.85 & 36.41 $\pm$ 7.02 & 7.17 $\pm$ 1.33 & 34.43 $\pm$ 6.30 \\
    
    S$^{18}$O (2,3 - 1,2) & < 0.14 & <  0.39 & < 0.42 & < 0.71 & 3.54 $\pm$ 0.72 & 2.29 $\pm$ 0.42 & < 0.79  & < 0.15  & 0.92 $\pm$ 0.17 \\
    
    $^{34}$SO (2,3 - 1,2) & 0.43 $\pm$ 0.09 & 3.41 $\pm$ 0.66 & < 0.42 & 1.78 $\pm$ 0.45 & 12.63 $\pm$ 2.53 & 9.72 $\pm$ 1.75 & < 0.75 & < 0.15  & 3.76 $\pm$ 0.69 \\
    
    SO$_2$ (3$_{1,3}$ - 2$_{0,2}$) & < 0.88 & 5.82 $\pm$ 1.63 & < 1.86 & 1.89 $\pm$ 2.88 & 11.35 $\pm$ 2.37 & 10.24 $\pm$ 1.92 & < 1.21 & < 0.77 & 6.33 $\pm$ 1.17 \\
    
    OCS (8 - 7) & < 1.55 & 20.57 $\pm$ 4.21 & < 3.37 & 32.54 $\pm$ 7.30 & 53.97 $\pm$ 10.96 & 33.44 $\pm$ 6.09 & < 8.31 & < 1.59 & 10.96 $\pm$ 2.07 \\
    
    $^{13}$CS (2 - 1) & 0.51 $\pm$ 0.10 & 1.58 $\pm$ 0.32 & 1.70 $\pm$ 0.35 & 3.44 $\pm$ 0.70 & 8.48 $\pm$ 1.70 & 1.06 $\pm$ 0.20 & 1.74 $\pm$ 0.37 & 0.20 $\pm$ 0.06 & 0.97 $\pm$ 0.18 \\
    
    C$^{34}$S (2 - 1) & 1.15 $\pm$ 0.20 & 5.41 $\pm$ 1.01 & 2.88 $\pm$ 0.55 & 8.83 $\pm$ 1.76 & 9.81 $\pm$ 1.99 & 3.71 $\pm$ 0.67 & 4.95 $\pm$ 0.96 & 0.70 $\pm$ 0.14 & 2.89 $\pm$ 0.53 \\

    CCS (7,8 - 6,7) & 6.47 $\pm$ 1.13 & 17.41 $\pm$ 3.26 & 11.41 $\pm$ 2.16 & 36.86 $\pm$ 7.33 & 28.56 $\pm$ 5.72 & 4.69 $\pm$ 0.85 & 17.83 $\pm$ 3.46 & 2.48 $\pm$ 0.47 & 11.95 $\pm$ 2.19 \\

    C$_3$S (16 - 15) & < 0.66  & < 2.49 & < 2.04  & < 1.90 & 3.08 $\pm$ 0.78 & < 1.19 & < 1.72 & < 0.59 & 1.70 $\pm$ 0.32 \\
    
    HCS$^{+}$ (2 - 1) & 1.17 $\pm$ 0.26 & 4.06 $\pm$ 0.82 & 4.06 $\pm$ 0.87 & 6.43 $\pm$ 4.25 & 5.89 $\pm$ 1.33 & 0.87 $\pm$ 1.00 & 3.79 $\pm$ 0.80 & -- & 2.41 $\pm$ 0.44 \\

    H$_2$CS (3$_{1,2}$ - 2$_{1,1}$) & 7.11 $\pm$ 1.26 & 28.54 $\pm$ 5.40 & 10.90 $\pm$ 2.23 & 43.59 $\pm$ 8.67 & 64.04 $\pm$ 12.82 & 15.36 $\pm$ 2.78 & 29.83 $\pm$ 5.77 & 3.72 $\pm$ 0.75 & 24.23 $\pm$ 4.43 \\
    
    HDCS (3$_{0,3}$ - 2$_{0,2}$) & 1.16 $\pm$ 0.22 & 6.20 $\pm$ 1.17 & 2.14 $\pm$ 0.44 & 7.14 $\pm$ 1.45 & 7.79 $\pm$ 1.59 & 3.21 $\pm$ 0.59 & 4.06 $\pm$ 0.86 & 0.58 $\pm$ 0.18 & 4.55 $\pm$ 0.83 \\ 
    \hline
    \end{tabular}
    }

\tablefoot{$^\text{a}$ SO (2,2 - 1,1) for L1495B; for the remaining cores the SO (2,3 - 1,2) transition is used.}

    \label{tab:4_abundances}
\end{table*}

\renewcommand{\arraystretch}{1.0}

\subsubsection{H$_2$ column density}

As the core evolves, the H$_2$ column density increases, making it a proxy for the evolutionary stage of starless and pre-stellar cores. Since molecular abundances are defined as the ratio between the molecular and the H$_2$ column density, these two quantities are not fully independent. Consequently, any observed trends in the H$_2$ column density reflect relative changes in molecular abundances with evolution rather than fully independent correlations. 

\begin{figure*}[h]
    \centering
    \includegraphics[width=\textwidth]{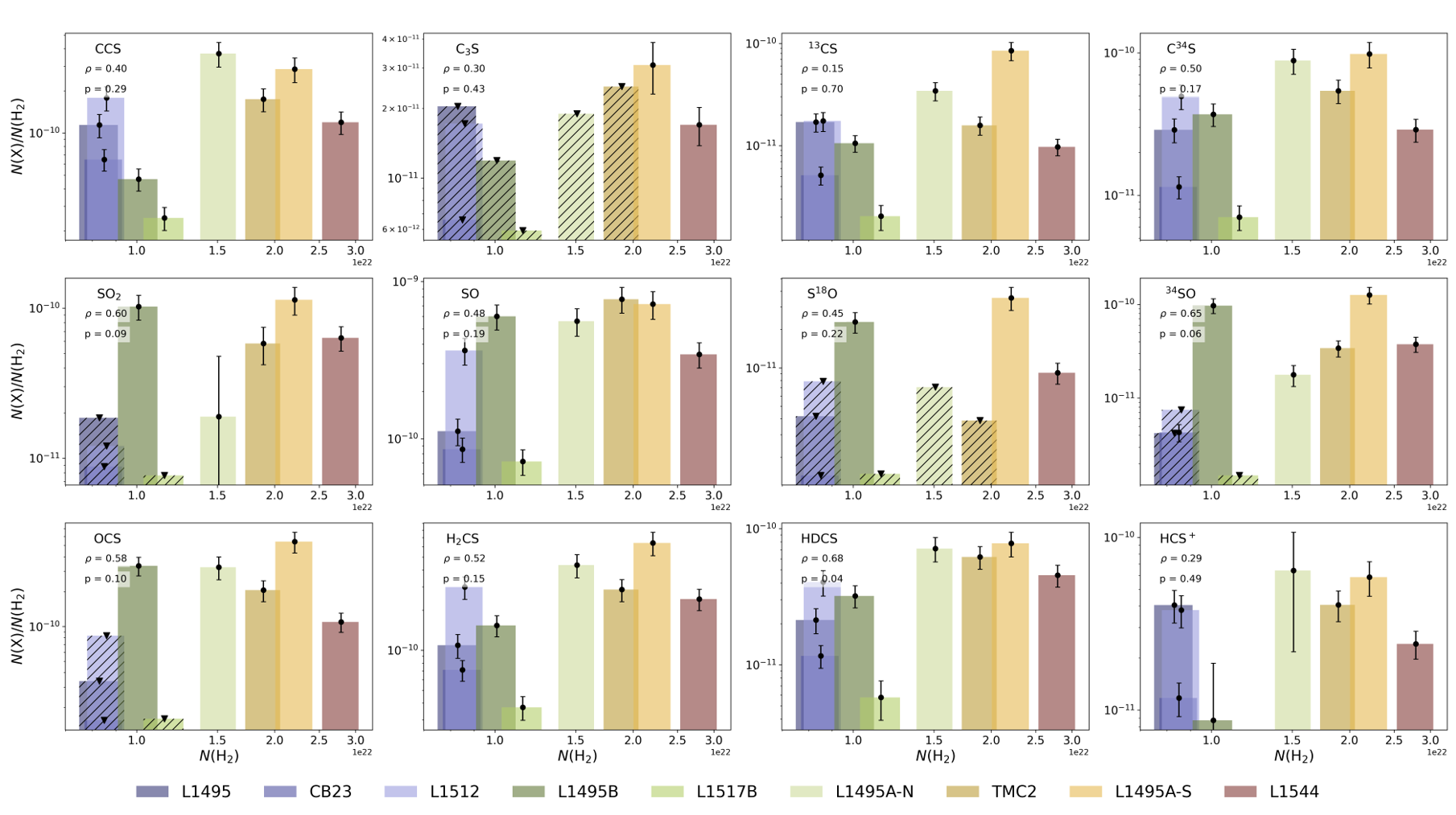}
    \caption{Molecular abundances as a function of the H$_2$ column density. Each panel shows one molecule, with bars representing the derived abundances for the different cores and widths corresponding to half the uncertainty in $N$(H$_2$). Hatched bars with a downward-facing triangle indicate upper limits (non-detections). Each panel also shows the Spearman rank correlation coefficient ($\rho$) and the corresponding $p$-value.}
    \label{fig:4_abundances_H2}
\end{figure*}

Figure \ref{fig:4_abundances_H2} shows the molecular abundances as a function of H$_2$ column density for all cores, with each subplot representing a molecule and hatched bars indicating upper limits. Overall, no clear trends are observed, as reflected in the Spearman rank correlation coefficient, which is generally around $\rho \sim 0.6$. We find the strongest positive correlation for $^{34}$SO with $\rho = 0.65$, which cannot be counted statistically significant. A leave-one-out analysis finds that excluding either L1495B or L1517B increases $\rho$ to $\sim 0.8$ ($p \sim 0.01$). A similar, but weaker behavior is seen for SO$_2$. HDCS also shows a possible correlation of $\rho = 0.79$ ($p = 0.02$), when L1517B is left out. While none of them is statistically significant, we find a slight positive correlation among $^{34}$SO, SO$_2$, and HDCS when L1517B is excluded.

Even though there are no clear monotonic trends, several molecules show similar abundance patterns across the cores. In particular, H$_2$CS and HDCS display a comparable relative behavior: the cores with a lower H$_2$ column density show more variation in abundance, whereas higher $N$(H$_2$) cores cluster more closely in their observed abundance. However, SO$_2$ and $^{34}$SO also show a similar pattern in their abundances across the cores. Besides these pairwise similarities, we also observe significant variation among the cores, with L1517B consistently showing the lowest abundances and L1495A-S and L1495A-N displaying the highest abundances for all molecules. 

We find that the cores with low $N$(H$_2$) values (L1495, CB23, L1512, and L1517B) display non-detections for most oxygen-bearing species, except SO. Only L1495B deviates from this trend. Even though it has a similar $N$(H$_2$) value, the core exhibits high abundances of oxygen-bearing species. Moreover, unlike the other cores embedded in the L1495 filament (L1495, L1495A-S, and L1495A-N), which exhibit similar abundances for all molecules, L1495B shows enhanced levels of oxygen-bearing sulfur molecules relative to carbon-bearing species. Given their proximity within the L1495 filament and their similar abundance levels, which agree within uncertainties for six out of ten molecules, it is likely that L1495A-N and L1495A-S have formed under similar physical and chemical conditions.

\subsubsection{N$_2$D$^+$/N$_2$H$^+$ ratio}

\begin{figure*}[h]
    \centering
    \includegraphics[width=\textwidth]{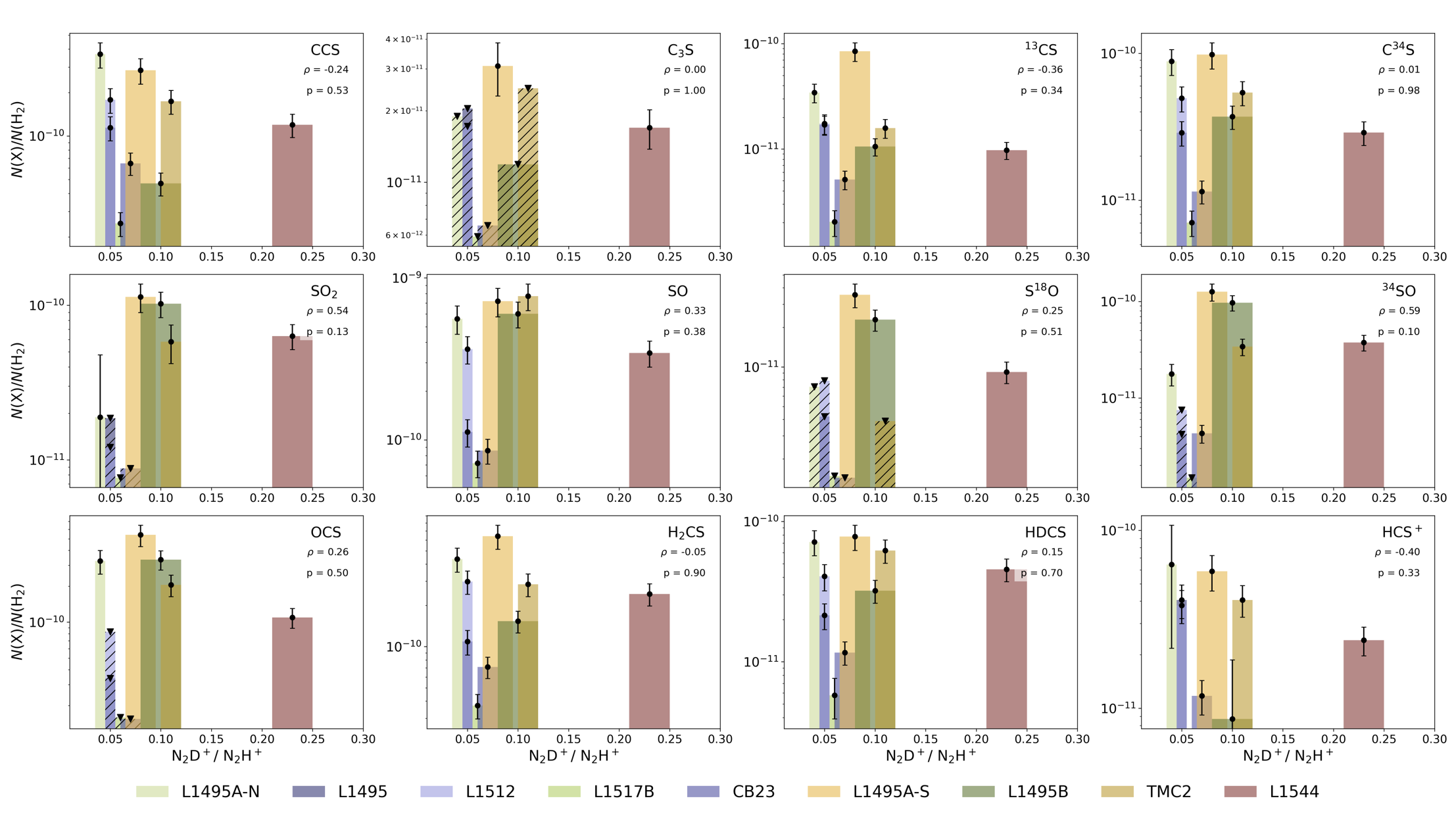}
    \caption{Same as Figure \ref{fig:4_abundances_H2} but for the N$_2$D$^+$/N$_2$H$^+$ ratio. Widths correspond to half the uncertainty in the N$_2$D$^+$/N$_2$H$^+$ ratio. }
    \label{fig:4_abundances_N2D+}
\end{figure*}

Due to the freeze-out of CO, the fraction of deuterated N$_2$H$^+$ increases in dense cores. Since N$_2$D$^+$ and N$_2$H$^+$ form through analogous reactions, the N$_2$D$^+$/N$_2$H$^+$ ratio (taken from \citealt{Crapsi2005}) serves as a good proxy for the core evolution. Figure \ref{fig:4_abundances_N2D+} shows the molecular abundances as a function of this ratio.

The  N$_2$D$^+$/N$_2$H$^+$ ratios  for all but one core are below 0.12. The exception, L1544, exhibits a ratio of 0.23, which is consistent with it being in a more advanced evolutionary state. Cores with lower N$_2$D$^+$/N$_2$H$^+$ ratios tend to have lower abundances of most oxygen-bearing species (except SO), similar to the behavior seen in the abundances as a function of the H$_2$ column density. This is true for L1495, L1512, and L1517B. L1495A-N does not follow the trend and exhibits high abundances for OCS and $^{34}$SO, despite having the lowest deuteration ratio. 

Overall, the scatter in abundances makes it difficult to identify any clear trends with N$_2$D$^+$/N$_2$H$^+$. A weak negative trend is visible only for HCS$^+$, but the low Spearman rank coefficient ($|\rho| < 0.6$ for all molecules) makes this behavior statistically insignificant.

\subsubsection{CO depletion factor}

\begin{figure*}[h]
    \centering
    \includegraphics[width=\textwidth]{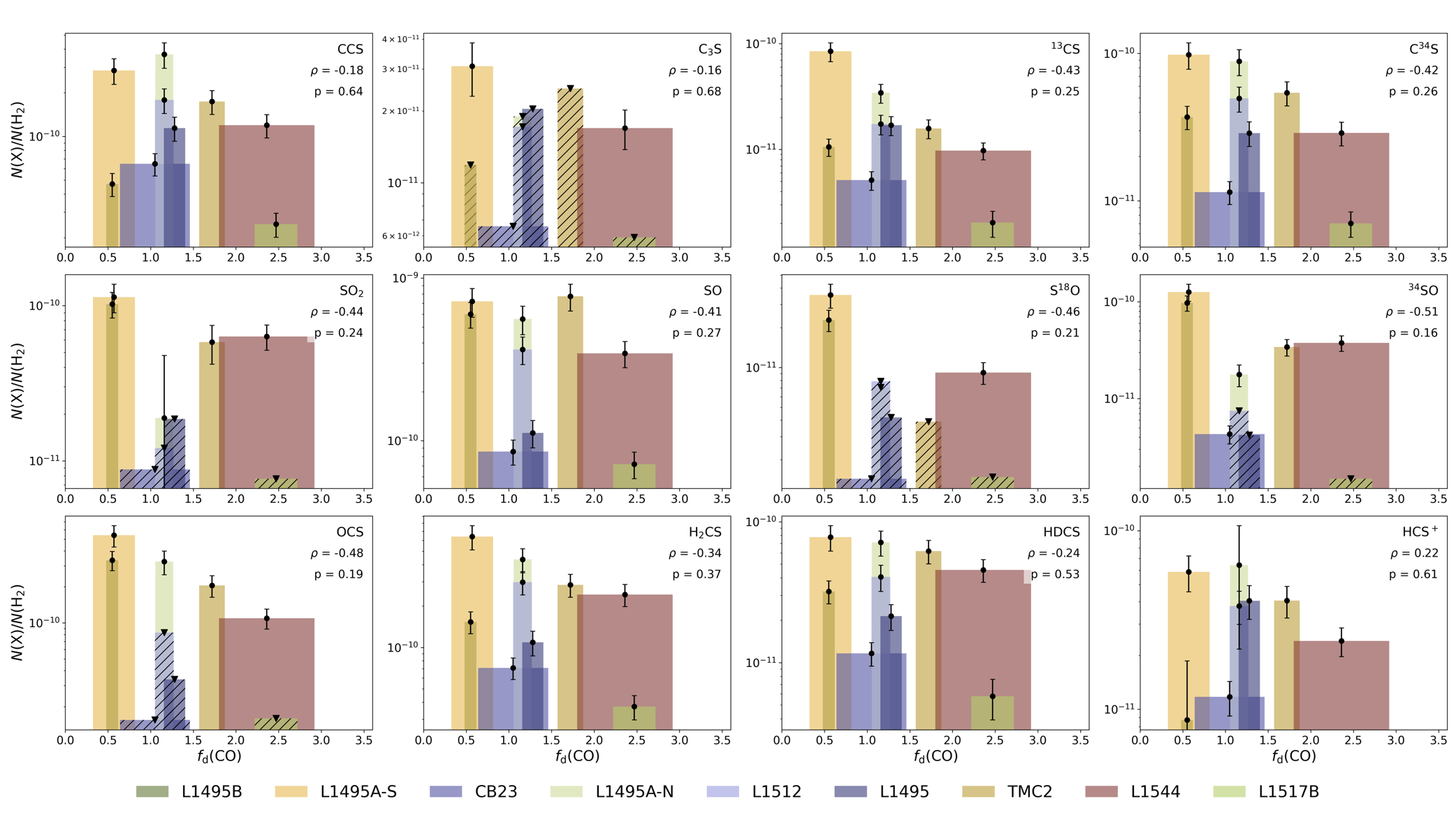}
    \caption{Same as Figure \ref{fig:4_abundances_H2} but for the CO depletion factor. Widths correspond to half the uncertainty in the CO depletion factor. }
    \label{fig:4_abundances_CO}
\end{figure*}

The chemical evolution of starless and pre-stellar cores can also be traced using the CO depletion factor, which provides a direct measure of CO freeze-out onto dust grains. Figure \ref{fig:4_abundances_CO} shows the molecular abundances as a function of the CO depletion factor. Similar to the other evolutionary tracers, no clear trend is visible. L1495A-S, one of the cores with high sulfur abundances, exhibits the lowest CO depletion factor, while L1517B has the highest CO depletion and the lowest abundance, despite not being classified as a highly evolved core (\citealt{Crapsi2005}). Within the L1495 cores, CO depletion and molecular abundances decrease gradually from L1495A-S to L1495A-N, then to L1495, with similar relative values. In contrast, L1495B deviates from this behavior and exhibits higher abundances of oxygen-bearing sulfur molecules compared to the other cores from the L1495 filament relative to the carbon- and hydrogen-bearing sulfur species.

The Spearman rank correlation coefficient shows no significant monotonic trend in the CO depletion factor. Nevertheless, several molecules show a decreasing trend with increasing CO depletion. This can be seen, for example, in the case of the CS isotopologs, $^{13}$CS and C$^{34}$S, or OCS (excluding non-detections). In contrast, the pure oxygen-bearing sulfur species, such as SO$_2$, SO, and its two isotopologs, exhibit a more complex behavior. A few molecules, such as H$_2$CS and C$^{34}$S, show two nearly parallel abundance sequences rather than a single evolutionary trend. Overall, these observations suggest weak negative trends with increasing CO depletion factor, but none are statistically significant.

\subsection{Ratios}
To probe various aspects of the chemistry in starless and pre-stellar cores, abundance ratios can be used. While some determine the dominant sulfur chemistry by comparing carbon- and oxygen-bearing molecules, others, such as the $^{34}$SO/SO ratio, trace isotopic fractionation. Here, we focused on six ratios, which we divided into three categories: ratios that trace isotopic fractionation (HDCS/H$_2$CS, $^{13}$CS/C$^{34}$S, and $^{34}$SO/SO), ratios that compare carbon and oxygen chemistry (CCS/ $^{34}$SO and C$^{34}$S/$^{34}$SO), and a ratio that probes the degree of carbon-chain growth within sulfur-bearing molecules (C$_3$S/CCS). The results are presented in Table \ref{tab:4_ratios} and are further analyzed as a function of three evolutionary tracers introduced in Sect. \ref{subsec:4_Abundances}. Note that for all cores, except L1544 and L1495A-S, upper limits were included in the ratio calculations, so the derived ratios are presented as upper or lower limits, depending on whether the non-detected molecule appears in the numerator or denominator. 

\renewcommand{\arraystretch}{1.3}
\begin{table*}[ht]
    \centering
    
    \caption{Observed abundance ratios for the core sample.}
    \resizebox{\textwidth}{!}{%
    \begin{tabular}{cccccccccc}
    \hline\hline
       Molecular Ratios&   CB23    &  TMC2 & L1495 & L1495A-N  & L1495A-S  & L1495B    & L1512 & L1517B    & L1544\\
    \hline
    HDCS/H$_2$CS & $0.16 \pm 0.04$ & $0.22 \pm 0.06$ & $0.2 \pm 0.06$ & $0.16 \pm 0.05$ & $0.12 \pm 0.03$ & $0.21 \pm 0.05$ & $0.14 \pm 0.04$ & $0.16 \pm 0.06$ & $0.19 \pm 0.05$ \\
    $^{34}$SO/SO & $0.05 \pm 0.01$ & $0.044 \pm 0.01$ & $< 0.038$ & $0.032 \pm 0.01$ & $0.18 \pm 0.05$ & $0.16 \pm 0.04$ & $< 0.021$ & $< 0.021$ & $0.11 \pm 0.03$ \\
    $^{13}$CS/C$^{34}$S & $0.45 \pm 0.1$ & $0.29 \pm 0.08$ & $0.59 \pm 0.2$ & $0.39 \pm 0.1$ & $0.86 \pm 0.2$ & $0.29 \pm 0.07$ & $0.35 \pm 0.1$ & $0.29 \pm 0.1$ & $0.34 \pm 0.09$ \\
    C$^{34}$S/$^{34}$SO & $2.66 \pm 0.73$ & $1.58 \pm 0.43$ & $> 6.84$ & $4.97 \pm 1.60$ & $0.78 \pm 0.22$ & $0.38 \pm 0.10$ & $> 6.61$ & $> 4.72$ & $0.77 \pm 0.20$ \\
    CCS/ $^{34}$SO & $15.01 \pm 4.11 $ & $5.11 \pm 1.38$ & $> 27.07$ & $20.75 \pm 6.65$ & $2.26 \pm 0.64$ & $0.48 \pm 0.12$ & $> 23.79$ & $> 16.61$ & $3.17 \pm 0.82$ \\
    C$_3$S/ CCS & $< 0.14$ & $< 0.23$ & $< 0.26$ & $< 0.077$ & $0.14 \pm 0.05$ & $< 0.37$ & $< 0.16$ & $< 0.26$ & $0.27 \pm 0.07$ \\
    \hline
    \end{tabular}
    }

    \label{tab:4_ratios}
\end{table*}

\subsubsection{Ratios that trace isotopic and deuterium fractionation}

\begin{figure*}[h]
    \centering
    \includegraphics[width=\textwidth]{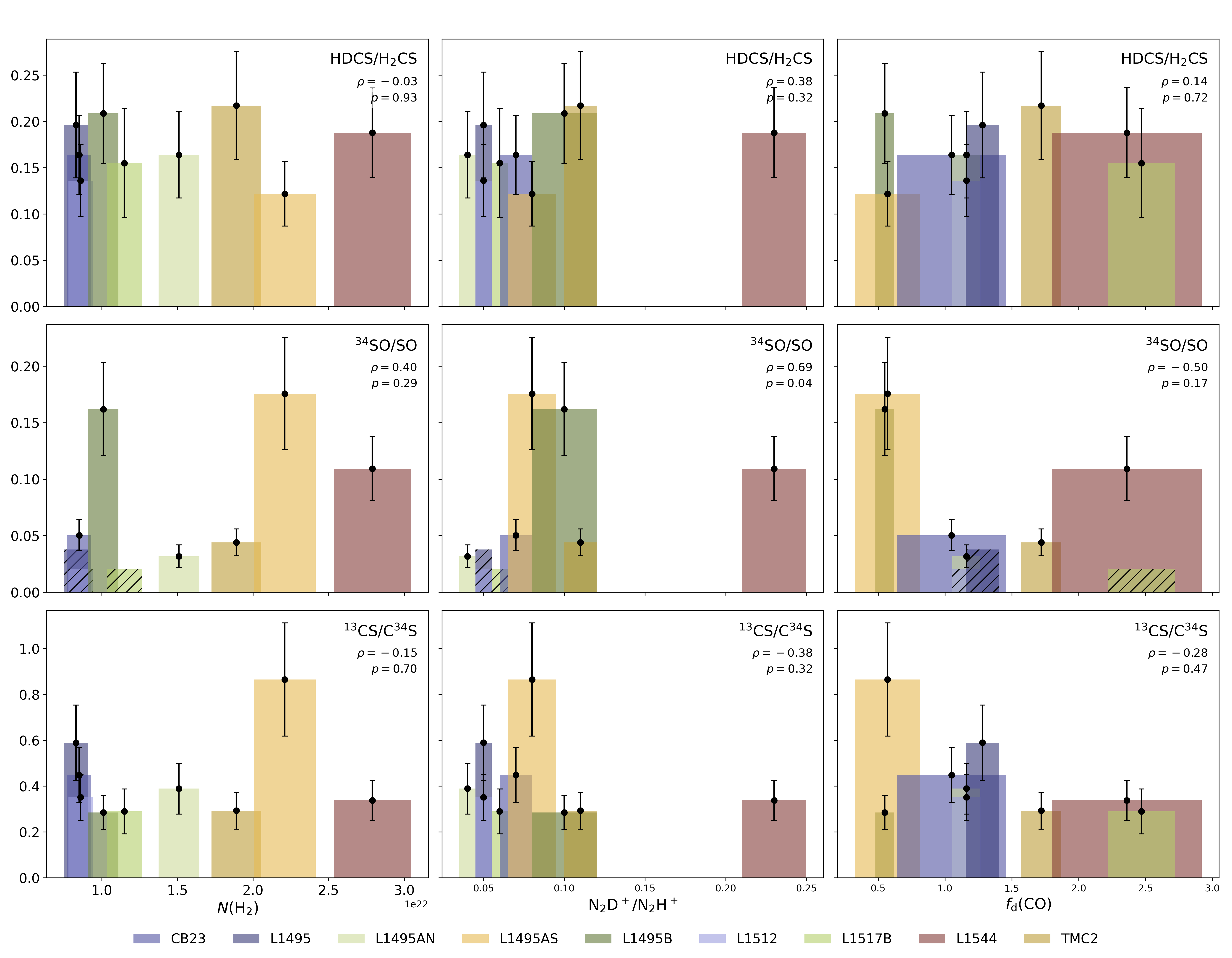}
    \caption{Molecular abundance ratios probing isotopic fractionation as a function of the three evolutionary tracers. For clarity, the width of each bar reflects half the uncertainty in the corresponding evolutionary tracer. Hatched bars indicate ratios where one of the two abundances represented an upper limit (non-detections). Each panel also shows the Spearman rank correlation coefficient ($\rho$) and the corresponding $p$-value.}
    \label{fig:4_ratios1}
\end{figure*}

Figure \ref{fig:4_ratios1} shows the isotopic ratios HDCS/H$_2$CS, $^{13}$CS/C$^{34}$S, and $^{34}$SO/SO as a function of the different evolutionary tracers. Among these, the HDCS/H$_2$CS column density ratio serves as a proxy for deuterium enrichment in the individual cores, and thus complements the N$_2$D$^+$/N$_2$H$^+$ ratio by tracing the deuteration within sulfur-bearing molecules. 

In the core sample, the deuteration ratios for each evolutionary tracer agree within uncertainties and span values between 0.12 and 0.21, without showing any monotonic trend. This is also reflected in the Spearman rank correlation coefficient, which shows a flat line for the H$_2$ column density and the CO depletion factor. 

The $^{34}$SO/SO ratio can indicate the optical depth of the SO line: if SO becomes optically thick, the ratio will be below the canonical isotopic value. In the optically thin limit, the expected  $^{34}$SO/SO ratio is $ 1/22 = 0.045$ (\citealt{WilsonWood1994}). Most sources are consistent with this value. However, L1495B, L1495A-S, and L1544 display particularly high ratios. For these cores, this may indicate either an elevated $^{34}$SO abundance or an underestimated SO abundance. Comparing the $^{34}$SO/SO ratio against the different evolutionary tracers, there is no notable trend. Excluding L1495B, the Spearman rank coefficient indicates a nonsignificant positive trend with the H$_2$ column density, while a slightly negative trend is found with the CO depletion factor. 

The idea behind the $^{13}$CS/C$^{34}$S ratio is similar to that of the $^{34}$SO/SO ratio. Taking the canonical values of $^{12}$C/$^{13}$C $= 68$ (\citealt{Milam2005}) and $^{34}$SO/SO $= 1/22 = 0.045$ (\citealt{WilsonWood1994}), we expect a ratio of $\sim 0.35$, assuming an optically thin line. All cores agree with the expected value within their uncertainties, except L1495A-S and L1495, which show higher ratios. Here, a possible explanation is an enhanced $^{13}$CS abundance or an underestimated C$^{34}$S abundance. When analyzed as a function of the different evolutionary tracers, the $^{13}$CS/C$^{34}$S ratio shows no notable trend. 

\subsubsection{Ratios that trace carbon- and oxygen-bearing sulfur species}

Carbon-chain chemistry is known as "early-type chemistry" in dense cores (\citealt{Suzuki1992}). As the core evolves, a larger fraction of carbon-bearing species is locked up in CO and frozen onto dust grains, reducing the gas-phase carbon and gradually slowing the formation of carbon chains. Consequently, oxygen-bearing molecules become relatively more abundant in the gas phase. Thus, comparing the carbon- and oxygen-bearing sulfur molecules, we would expect a decrease in both the ratios of C$^{34}$S/$^{34}$SO and CCS/$^{34}$SO as a function of the evolutionary tracer. While CCS/$^{34}$SO focuses more on the carbon chain chemistry, C$^{34}$S/$^{34}$SO traces the more general balance between the carbon- and oxygen-bearing sulfur molecules. 

\begin{figure*}[h]
    \centering
    \includegraphics[width=\textwidth]{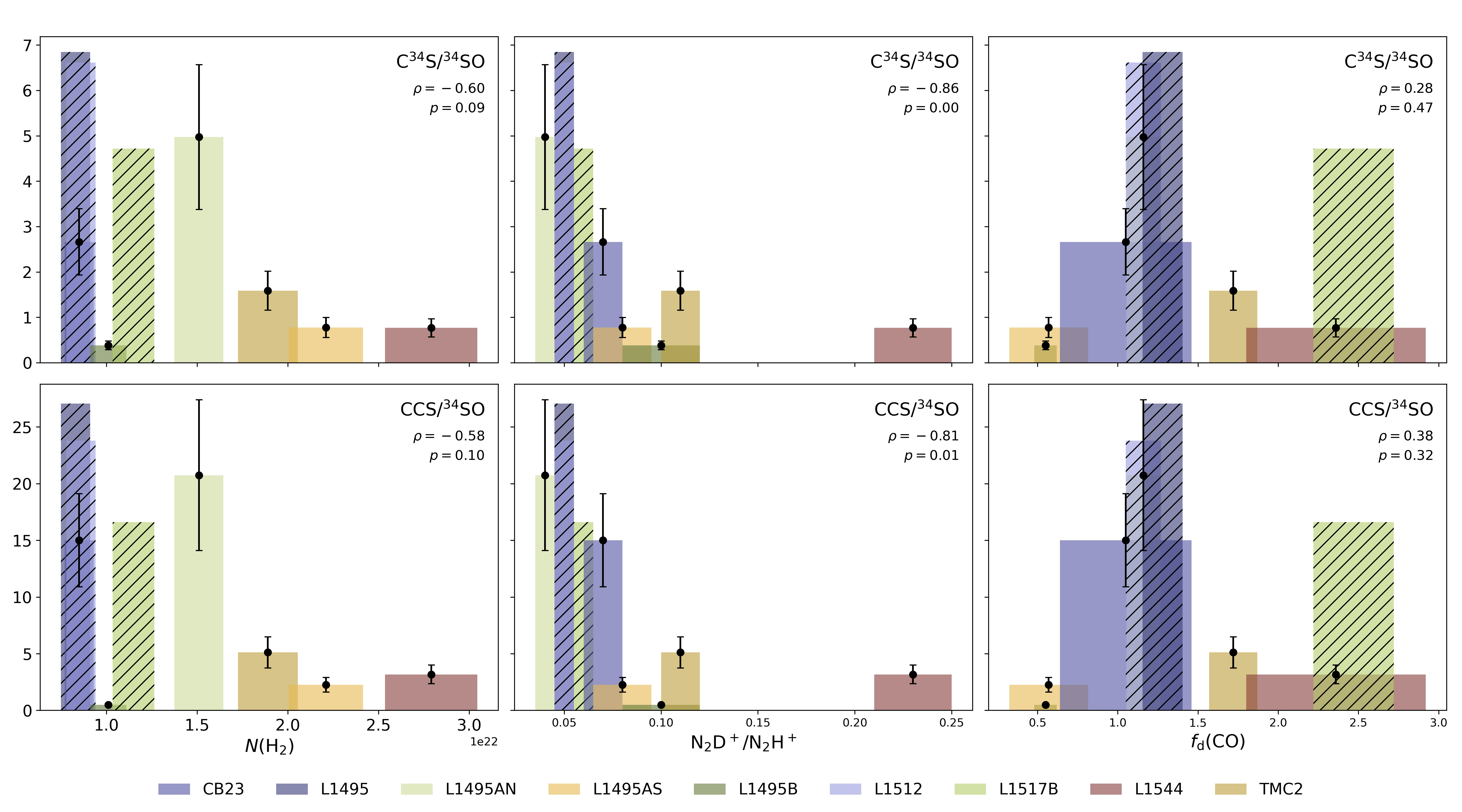}
    \caption{Same as Figure \ref{fig:4_ratios1} but for the carbon- and oxygen-bearing sulfur species.}
    \label{fig:4_ratios2}
\end{figure*}

Figure \ref{fig:4_ratios2} shows both ratios as a function of the H$_2$ column density, the N$_2$D$^+$/N$_2$H$^+$ ratio, and the CO depletion factor. For the CCS/$^{34}$SO ratio, a moderate negative trend is found with the H$_2$ column density ($\rho = -0.58$, $p = 0.10$). L1495B strongly influences this trend, as it shows a comparatively high abundance across all oxygen-bearing sulfur molecules. When excluding L1495B, the trend becomes more pronounced and yields a statistically significant anticorrelation ($\rho = -0.81$, $p = 0.01$). The Spearman rank coefficient shows a similar value for the N$_2$D$^+$/N$_2$H$^+$ ratio with $\rho = -0.81$, $p = 0.01$, which does not change when removing individual cores from the sample. For the CO depletion factor, no correlation was found.

The C$^{34}$S/$^{34}$SO ratio exhibits a similar behavior as the CCS/$^{34}$SO ratio: for the H$_2$ column density, a moderate anticorrelation is found ($\rho = -0.60$, $p = 0.09$). This correlation becomes more pronounced if L1495B is excluded from the analysis. For the N$_2$D$^+$/N$_2$H$^+$ ratio, the anticorrelation is even stronger ($\rho = -0.86$, $p < 0.01$) and remains mostly unchanged when individual cores are removed from the analysis. Again, no clear trend is observed with increasing CO depletion factor.

\subsubsection{Carbon chains}

The C$_3$S/CCS ratio provides information on the length of the carbon chains in the core. While carbon chemistry is generally more prominent in the earlier stages, C$_3$S forms, among other pathways via the surface reaction CCS + C (\citealt{LaasCaselli2019}). Based on the column densities reported by \citet{Vastel2018} for L1544, we would expect a ratio of 0.2-0.3. This compares well to the derived ratio of 0.27 for the same core. For L1495A-S, the value is 0.14, which agrees with the idea that it is less evolved. For the remaining cores, C$_3$S is not detected.

\begin{figure*}[h]
    \centering
    \includegraphics[width=\textwidth]{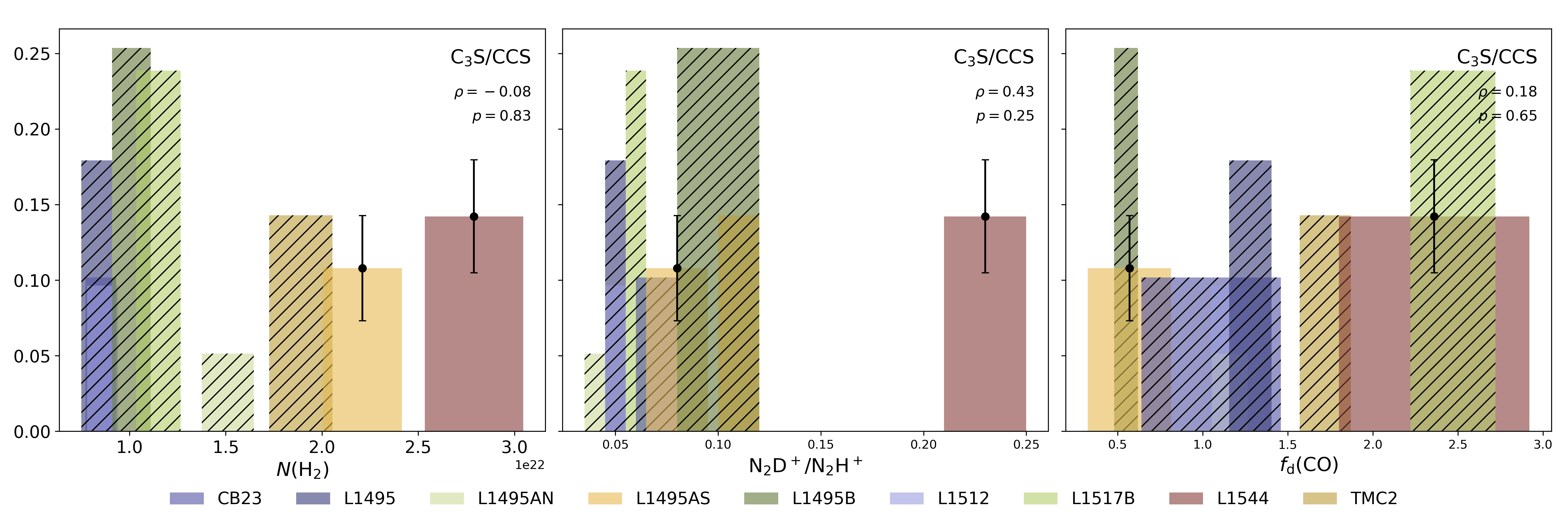}
    \caption{Same as Figure \ref{fig:4_ratios1} but for the C$_3$S/CCS ratio.}
    \label{fig:4_ratios3}
\end{figure*}

Figure \ref{fig:4_ratios3} shows the ratio as a function of evolutionary tracers. We observe an increase as a function of the CO depletion factor and the N$_2$D$^+$/N$_2$H$^+$ ratio, as suggested by the Spearman rank correlation coefficient. In contrast, for the H$_2$ column density, no trend is visible. Given the large number of upper limits, this trend must be interpreted with caution. 

\section{Single-point chemical models}
\label{sec:Models}
To investigate whether our current knowledge of sulfur chemistry can explain the observed abundances and the lack of clear evolutionary trends, we employed gas-grain chemical simulations with the code \texttt{pyRate}, varying the initial sulfur abundance ($X_\text{i}$(S)) and identifying the best-fitting model for each core. The specific details of the model, including the chemical network, are described in \citet{Olli2015} and \citet{Olli2019} and are briefly summarized here. Note that the network also incorporates the sulfur reactions reported in \citet{LaasCaselli2019}. We applied a 0D chemical model assuming a constant temperature of $T_\text{gas} = T_\text{dust}$ = \SI{10}{K}, a density of $n$(H$_2) = 10^4$ cm$^{-3}$, and a cosmic ray ionization rate of $\zeta_2 = 1.3 \cdot 10^{-17}$ s$^{-1}$. These values were chosen to represent the typical conditions in the outer envelopes of starless and pre-stellar cores, which are known to host sulfur-bearing species. We adopted three different values for $X_\text{i}$(S)$ = 8\cdot 10^{-6}$, $8\cdot 10^{-7}$, and $8\cdot 10^{-8}$, representing an un-depleted sulfur abundance and increasing degrees of depletion. The three values of $X_\text{i}$(S) are often referred to as high, intermediate, and low $X_\text{i}$(S) in the following text.

\subsection{Comparison to models}
We evaluated the detected sulfur-bearing molecules at a chemical age of $10^6$ years, an approximate timescale for dense cores (\citealt{Evans2009}). Isotopes of sulfur, oxygen, and carbon are not included in the chemical network. Thus, HDCS was the only isotopolog considered in the models. Since CS was previously excluded due to a lack of data caused by optically thick effects, we inferred its observational abundance from its optically thin isotopologs. For cores where CS and both $^{13}$CS and C$^{34}$S were detected, the CS/$^{13}$CS and CS/C$^{34}$S ratios were calculated and compared to the canonical elemental ratios in the local ISM of $^{12}$C/$^{13}$C = 68 (\citealt{Milam2005}) and $^{32}$S/$^{34}$S = 22 (\citealt{WilsonWood1994}). On average, both ratios are around half of the canonical values. Since the $^{13}$CS line in CB23 exhibits a broader linewidth than the average of the core, and as the CS/C$^{34}$S value is better represented, we adopted C$^{34}$S together with the canonical isotopic ratio to determine the CS abundance. We divided the nine remaining sulfur-bearing molecules into three groups: O-bearing sulfur molecules (SO, SO$_2$, and OCS), C-bearing sulfur molecules (CS, CCS, and C$_3$S), and HC-bearing sulfur molecules (HCS$^+$, H$_2$CS, and HDCS). Note that for this analysis, no upper limits were considered. 

\begin{figure*}[h]
    \centering
    \includegraphics[width=\textwidth]{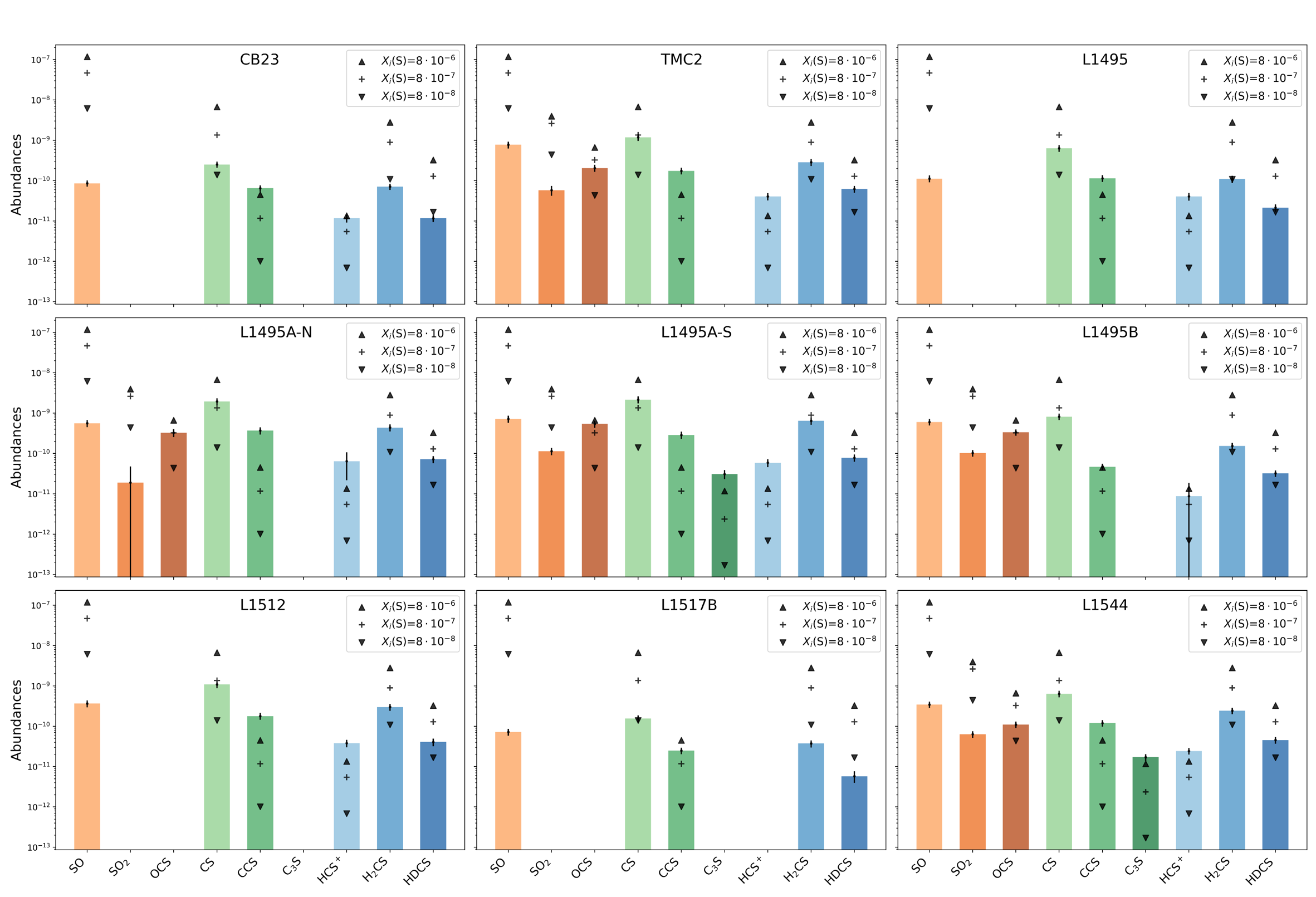}
    \caption{Comparison between the observational data and the model abundance at three different $X_\text{i}$(S). The colored bars represent the observations, while the different symbols show the models evaluated at $10^6$ years.}
    \label{fig:5_comp_models_abundances}
\end{figure*}

Figure \ref{fig:5_comp_models_abundances} compares the observational abundances with the models at the three different $X_\text{i}$(S), evaluated at 10$^6$ years. Molecules of the same group are presented in similar colors for better distinction. For oxygen-bearing molecules, the models generally over-predict the observed abundances, with SO and SO$_2$ differing by up to an order of magnitude across the cores. OCS is better reproduced and best represented by the intermediate $X_\text{i}$(S). In contrast, the carbon-bearing molecules CCS and C$_3$S are mostly under-predicted and thus best match the high $X_\text{i}$(S) model. CS acts as an outlier within the carbon group, with the best-fitting model varying between intermediate and high $X_\text{i}$(S). For the hydrogen-bearing species, HCS$^+$ is generally under-predicted, whereas H$_2$CS and HDCS are well reproduced. In all cores, both H$_2$CS and HDCS agree with the same model, indicating that the models successfully reproduce the observed HDCS/H$_2$CS ratios.

\subsection{Mean distance of disagreement}
To determine which model best agrees with the observed abundances, we employed the mean distance of disagreement ($D(t)$) approach introduced in \citet{IqbalWakelam2018} and incorporated a method that accounts for observational uncertainties (\citealt{Wakelam2006}). We implemented this via
\begin{equation}
    D(t) = \frac{1}{N_\text{obs} } \sum_i |d_\text{i} | ,
\end{equation}
where $d_\text{i}$ describes the distance between the observational and model abundances and $N_\text{obs}$ is the number of points taken into account. The individual distance element ($d_i$) is calculated differently depending on whether the observed abundance is higher or lower than the model value. We used $d_\text{i} = \log (X_\text{obs, min}) - \log (X_\text{model})$ for the case $X_\text{obs}$>$X_\text{model}$ and $d_\text{i} = \log (X_\text{model}) - \log (X_\text{obs, max})$ for the case $X_\text{obs}$ < $X_\text{model}$. 
If the observational and model abundances agree, $d_\text{i} = 0$. The smaller the resulting number, the greater the agreement between the model and the data. 

Table \ref{tab:5_mdd_results} represents the mean distance of disagreement for the individual cores evaluated at $10^6$ years with different $X_\text{i}$(S). For six cores, the best model (indicated by the bolded value in Table \ref{tab:5_mdd_results}) is the one with intermediate $X_\text{i}$(S). The remaining three fit best with the depleted (low) $X_\text{i}$(S). Regarding the individual groups, the results are similar to those shown in Figure \ref{fig:5_comp_models_abundances}: while the O-bearing group matches the lowest $X_\text{i}$(S), the carbon-bearing group generally agrees best with the highest, except for L1517B, where the intermediate $X_\text{i}$(S) reproduces the observations the best. Only the HC-bearing group is consistent with different $X_\text{i}$(S) across the cores. For this group, the model that best reproduces the observations is generally the one with the lowest mean distance of disagreement. The only exception is L1495B. While the HC-bearing group matches the lowest $X_\text{i}$(S), considering all molecules, the intermediate sulfur value fits best according to the mean distance of disagreement. 

\begin{table}[h]
    \centering
    \caption{Results of the mean distance of disagreement analysis, including all the detected molecules in the individual cores.}
    \scalebox{0.9}{%
    \begin{tabular}{cccc}
    \hline\hline
    Core & $X_\text{i}$(S)$ = 8 \cdot 10^{-6}$ & $X_\text{i}$(S) $ = 8 \cdot 10^{-7}$ & $X_\text{i}$(S)$ = 8 \cdot 10^{-8}$ \\
    \hline
    CB23    &  1.317 & 1.146 & \textbf{0.955}  \\
    TMC2    &  0.833 &\textbf{ 0.744} & 1.067  \\
    L1495   &  1.190 & 1.098 & \textbf{1.029}  \\
    L1495A-N&  0.801 & \textbf{0.753} & 1.180  \\
    L1495A-S&  0.726 & \textbf{0.725} & 1.209  \\
    L1495B  &  0.775 & \textbf{0.615} & 0.718  \\
    L1512   &  1.008 & \textbf{0.917} & 1.111  \\
    L1517B  &  4.644 & 4.332 & \textbf{3.899}  \\
    L1544   &  0.878 & \textbf{0.787} & 0.945  \\
    \hline
    \end{tabular}
    }
    \tablefoot{In bold, the lowest value representing the best-fitting model.}
    \label{tab:5_mdd_results}
\end{table}

\section{Discussion}
\label{sec:Discussion}

We observe clear variations in the molecular abundances across the cores, yet no statistically significant trends with the three evolutionary tracers. L1495B clearly stands out for its high abundances of oxygen-bearing sulfur species. Conversely, L1517B exhibits low abundances and a high CO depletion factor, even though it was classified as a starless core (\citealt{Crapsi2005}). This might be due to the dense and quiescent conditions under which the core formed, rather than an advanced evolutionary stage. Additionally, we find that ratios of carbon- and oxygen-bearing molecules decrease with increasing H$_2$ column density and N$_2$D$^+$/N$_2$H$^+$, but show no notable trend when compared to the CO depletion factor. This suggests that the environment or the location of the molecular emission strongly influences the observed chemistry.

Indeed, most sulfur-bearing molecules have their emission peak not at the dust continuum maxima, but rather in the outer envelope of the cores (e.g., \citealt{Spezzano2017}; \citealt{Tafalla2006Sulfurouterlayer}). As the outer layers are less tightly coupled to the evolutionary state of the core's center, it might explain why no clear trends between the sulfur abundances and the adopted evolutionary tracers were observed. Conversely, the outer envelopes are more directly affected by their environment, which can explain the variation in the observed abundances between the cores (\citealt{Fuente2023}).

Several observational surveys have cataloged young stellar objects and energetic sources in the Taurus Molecular Cloud and thus provide additional context on the environments of the different filaments. These include, for example, catalogs of young stars from the \textit{Spitzer Space Telescope} (\citealt{Rebull2010}) and the \textit{XMM-Newton} Extended Survey of the Taurus molecular cloud (XEST; \citealt{Guedel2007}), which identified X-ray emitting sources. Notably, the XEST survey locates the strongest X-ray source of the Taurus Molecular cloud, V410 Tau ABC (23-032), at a projected distance of 0.01 pc from L1495A-N (assuming a distance of 129.9 pc for the L1495 filament; \citealt{Galli2009}). At such a close projected distance, X-rays can locally increase the ionization rate and the cascade of ion-molecule chemistry. Under favorable conditions, it elevates the abundance levels of H$_3^+$, N$_2$H$^+$, and HCS$^+$ while also contributing to the destruction of CO (\citealt{Staeuber2005}). To evaluate the potential impact on our cores, we estimated the X-ray ionization rate produced by V410 Tau at the positions of the two closest cores, L1495A-N and L1495A-S, by integrating the attenuated stellar X-ray photon flux over energy using photo-absorption cross sections and accounting for secondary electron ionizations (\citealt{Dalgarno1999}). The resulting ionization rates, summarized in Table \ref{tab:6_zeta_Xray} for different assumed $N$(H$_2$) values between source and core, indicate that for a typical column density of $N_{\mathrm H}=10^{22}\,\mathrm{cm^{-2}}$, X-rays clearly dominate the ionization in the vicinity of L1495A-N, while their effect becomes negligible at the larger distance of L1495A-S.

Interestingly, L1495A-N shows the highest abundances of several sulfur-bearing molecules in our sample, including HCS$^+$, CCS, C$^{34}$S, H$_2$CS, and HDCS. Motivated by the strong X-ray influence of V410 Tau on L1495A-N, we increased the ionization rate in our chemical models by one and two orders of magnitude to test whether this improves the agreement with the observed abundances across the cores. Assuming the low $X$(S$_\text{i}$), increasing the ionization rate by one order of magnitude improves the overall agreement between modeled and observed abundances for all cores, which may reflect that the different molecules trace different layers of gas within the cores. The effect differs for the different sulfur species: oxygen-bearing sulfur molecules are generally better reproduced with the standard ionization rate, whereas carbon-bearing sulfur species show better agreement with the enhanced rate. The HC-group behaves differently with HCS$^+$ better reproduced at the one order of magnitude higher ionization, while H$_2$CS and HDCS depend on the core. However, at higher ionization rates, the models no longer reproduce the observed D/H ratios, which were well reproduced by the standard rate. The differences between the standard and enhanced ionization models are particularly high for L1495A-N and L1495A-S, where the higher ionization rate provides a better fit. In contrast, increasing the standard ionization rate by two orders of magnitude leads to deterioration between the model and observations, as all molecules in all cores are systematically under-predicted, particularly SO$_2$ and OCS.

\begin{table}
    \centering
    \caption{Estimated X-ray ionization rates from V410 Tau for L1495A-N and L1495A-S for different intervening hydrogen column densities.}
    \begin{tabular}{ccc}
    \hline\hline
    $N$(H$_2$) & $\zeta_{\mathrm X}$ (L1495A-N) & $\zeta_{\mathrm X}$ (L1495A-S) \\
    (cm$^{-2}$)  & (s$^{-1}$) & (s$^{-1}$) \\
    \hline
    0 & $2.8\times10^{-15}$ & $1.2\times10^{-17}$ \\
    $10^{21}$ & $1.3\times10^{-15}$ & $5.4\times10^{-18}$ \\
    $10^{22}$ & $1.6\times10^{-16}$ & $6.7\times10^{-19}$ \\
    $10^{23}$ & $1.3\times10^{-17}$ & $5.4\times10^{-20}$ \\
    \hline
    \end{tabular}
    \label{tab:6_zeta_Xray}
\end{table}

In addition to X-ray sources, nearby protostars can impact the sulfur chemistry through heating, shocks from accretion/outflows, or UV radiation (e.g., \citealt{Codella2025}). Using the catalog presented in \citet{Rebull2010}, we find that the cores L1512, TMC2, and L1517B are relatively isolated, with no protostar within a projected distance of $\sim$ 0.25 pc. In contrast, the L1495 filament, especially the northern part, where the cores of this sample are located, contains many protostars. This part has been identified as one of the most evolved substructures within the filament, due to its high number of protostars (\citealt{Hacar2013}). For most of the starless cores in this filament, one or more protostars are found within a projected radius of 0.2 pc, except for L1495B, which is more isolated (the closest protostar is at $\sim$ 0.31 pc). The proximity of young stars may therefore contribute to the enhanced sulfur abundance levels in L1495A-N and L1495A-S.

Because L1495B is more shielded from external UV radiation, it exhibits high abundances of oxygen-bearing sulfur species. A higher degree of shielding favors the conversion of atomic carbon into CO, thus reducing the availability of free carbon. At the same time, enough atomic oxygen remains to sustain oxygen-driven reactions, thus allowing molecules such as SO or SO$_2$ to form more efficiently (\citealt{Lattanzi2020}). The L1495 cores thus demonstrate that local conditions can cause differences in sulfur abundance, even though they are separated by only 0.3 pc.

Within the core sample, L1517B stands out for its relatively low abundances of sulfur-bearing species and its high CO depletion factor. \citet{Hacar2011} has studied the L1517 region in further detail and found it to be quite quiescent and dense, with no embedded protostars. The gas surrounding the starless cores exhibits subsonic velocity fields, indicating the absence of strong shocks. The generally low abundance of sulfur-bearing molecules, as well as the numerous non-detections of oxygen-bearing sulfur species in L1517B, support the idea of a quiet environment. The high CO depletion factor and the low N2D+/N2H+ ratio are likely a consequence of physical and chemical evolution not always progressing simultaneously and being influenced by local environmental effects. Therefore, the CO depletion factor alone does not indicate that this core is more evolved than the other starless cores in the sample.

Regarding the molecular ratios, the two ratios comparing carbon and oxygen chemistry show a dependence on two of the three evolutionary parameters: the H$_2$ column density and the N$_2$D$^+$/N$_2$H$^+$ ratio. As these evolutionary tracers increase, the ratios decrease. This agrees with the general trend that carbon chemistry dominates in less evolved cores, while oxygen chemistry becomes more important at later stages and is further confirmed by the non-detections in oxygen-bearing sulfur species for cores with low $N$(H$_2$) and N$_2$D$^+$/N$_2$H$^+$. A possible reason the ratios comparing carbon- and oxygen-bearing sulfur species do not correlate with the CO depletion factor could be that these ratios are more sensitive to specific chemical pathways that do not strictly trace the CO depletion factor.

The chemical models considered density and temperature to represent the core's envelope and showed variations in the best-fitting $X_\text{i}$(S) across the molecules. While OCS is reproduced within an order of magnitude for the best fitting $X_\text{i}$(S), other species, such as SO, SO$_2$, and C$_3$S, are consistently over- or under-predicted regardless of the model. Additional tests, in which the time step for extracting the modeled abundances was reduced to $10^5$ years (before 10$^6$ years), showed a change in behavior: the oxygen-bearing species better agree with the models, whereas the carbon-bearing molecules are over-predicted. This indicates that simply adjusting the timescale cannot reproduce both the carbon- and oxygen-bearing sulfur molecules simultaneously.

A similar behavior was found in \citet{Vastel2018}, who found that no single value of the elemental sulfur abundance matched the observations of 21 sulfur-bearing molecules simultaneously in the pre-stellar core L1544. When varying the time in the models, we find that carbon- and oxygen-bearing sulfur species are reproduced differently, indicating that no single chemical age can simultaneously reproduce the abundances of all observed molecules. An alternative explanation is that the assumed C/O ratio in the models does not accurately describe the chemical conditions in the envelopes of dense cores. \citet{Byrne2026} show that even slight variations in the C/O ratio can strongly change the relative abundances of carbon- and oxygen-bearing molecules. Such sensitivities imply that simplified model assumptions might not be sufficient. Instead of using 0D chemical models, a 1D or higher-dimensional model would likely better capture the core's physical structure and thus provide more accurate results, as seen for example in  \citet{Jensen2023}, \citet{Priestley2023}, and \citet{Priestley2024}. Given the limited observational constraints on the formation pathways of sulfur-bearing species, such as non-detections or low S/R, our results further indicate that our current understanding of sulfur chemistry is incomplete.

\section{Conclusion}
\label{sec:Conclusion}

In this work we have presented single-point observations of 13 different sulfur-bearing molecules toward seven starless and two pre-stellar cores in the Taurus Molecular Cloud. By comparing the abundances and selected molecular ratios with three evolutionary tracers, we tested how sulfur chemistry behaves across cores at different evolutionary stages. Additionally, the CO-depletion factors for CB23, L1495A-N, L1495A-S, L1495B, and L1512 were determined for the first time. 

The observed abundances show notable variations between the cores and display weak or absent trends with respect to three evolutionary tracers, $N$(H$_2$), N$_2$D$^+$/N$_2$H$^+$, and the CO depletion factor. This agrees with the idea that sulfur-bearing molecules mainly trace the outer envelopes of the cores rather than the central regions, which are better represented by these tracers. Two molecular ratios that probe the difference between carbon- and oxygen-bearing sulfur species, C$^{34}$S/$^{34}$SO and CCS/$^{34}$SO, exhibit a more pronounced dependence on two of the three evolutionary indicators. This indicates that, while individual sulfur-bearing species are weak tracers of evolution, the relative abundances of carbon- and oxygen-bearing sulfur molecules are sensitive to both physical and chemical conditions. Additionally, we ran 0D chemical models with varying initial sulfur abundances. At $10^6$ years, the models reproduce the observed abundances for some species, such as OCS, H$_2CS$, and HDCS, within an order of magnitude, but cannot replicate the abundances of all sulfur-bearing molecules simultaneously. In particular, the oxygen-bearing molecules SO and SO$_2$ are consistently over-predicted, while carbon-bearing species, such as CCS and C$_3$S, are generally under-predicted. This difference between the carbon- and oxygen-bearing becomes even more prominent when varying the chemical age, reinforcing that, with the current 0D chemical models, no single chemical age can simultaneously describe all the observed sulfur-bearing species.

The results indicate that no single evolutionary parameter can describe the sulfur chemistry in starless and pre-stellar cores. Instead, the chemistry is influenced by a combination of the chemical age and different local environmental conditions. We find that nearby protostars or X-ray sources can measurably affect the detected sulfur abundance levels, leading to clear differences in abundance levels between sources. A single evolutionary tracer or model parameter cannot capture these variations.

The inability of 0D chemical models to simultaneously reproduce the observed sulfur abundances of all detected molecules demonstrates both their limitations and the complexity of sulfur chemistry. While the models can capture general chemical trends, the 0D structure limits the performance. It thus fails to account for variations in the physical conditions of individual cores or the influence of the local environment. Additionally, several formation and destruction pathways of sulfur-bearing molecules remain unknown under cold-core conditions. Together, this suggests that the sulfur chemistry in dense cores is a combination of the physical structure, environmental effects, and incomplete chemical knowledge. Future studies should thus focus on targeted observations of individual cores in well-characterized regions, such as L1517, combined with more complex modeling approaches to obtain a more comprehensive picture of sulfur chemistry in dense cores. 

\section*{Data availability} Table B.1 is only available in electronic form at the CDS via anonymous ftp to \url{cdsarc.u-strasbg.fr} (130.79.128.5) or via \url{http://cdsweb.u-strasbg.fr/cgi-bin/qcat?J/A+A/}.\\

\begin{acknowledgements} The authors wish to thank the anonymous referee for their comments, which helped to improve the clarity and quality of this work. L.S., S.S., and S.S.J. wish to thank the Max Planck Society for the Max Planck Research Group funding.\end{acknowledgements}

\bibliographystyle{aa}
\bibliography{bibliography}

\begin{appendix}
\label{appendix}
\onecolumn

\section{Plots of observed spectra}
\label{sec:AppendixA}

\begin{figure*}[h!]
    \centering
    \includegraphics[width=0.95\textwidth]{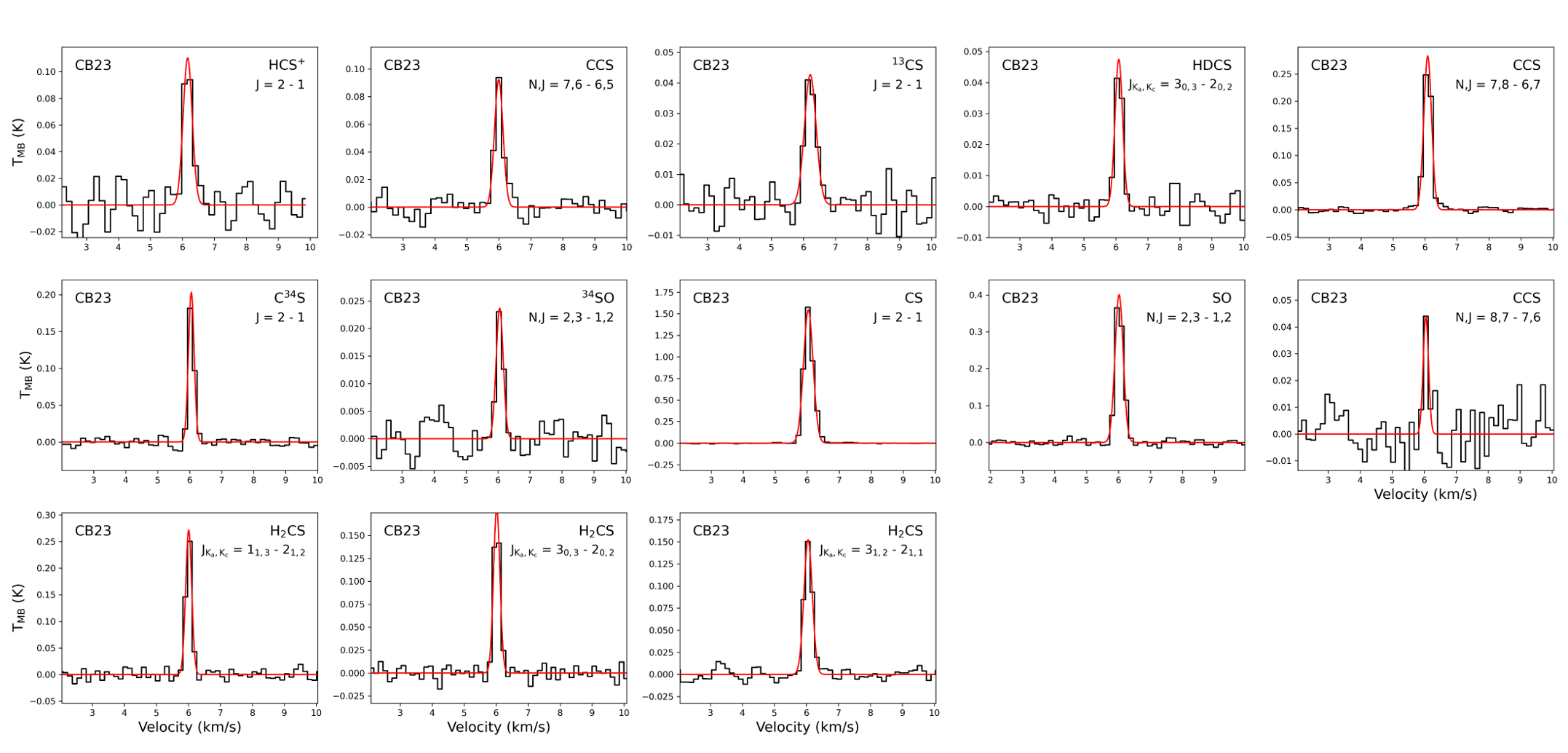}
    \caption{Same as Figure \ref{fig:3_linesL1495AS} but for the starless core CB23. }
    \label{fig:A_linesCB23}
\end{figure*}

\begin{figure*}[h!]
    \centering
    \includegraphics[width=0.95\textwidth]{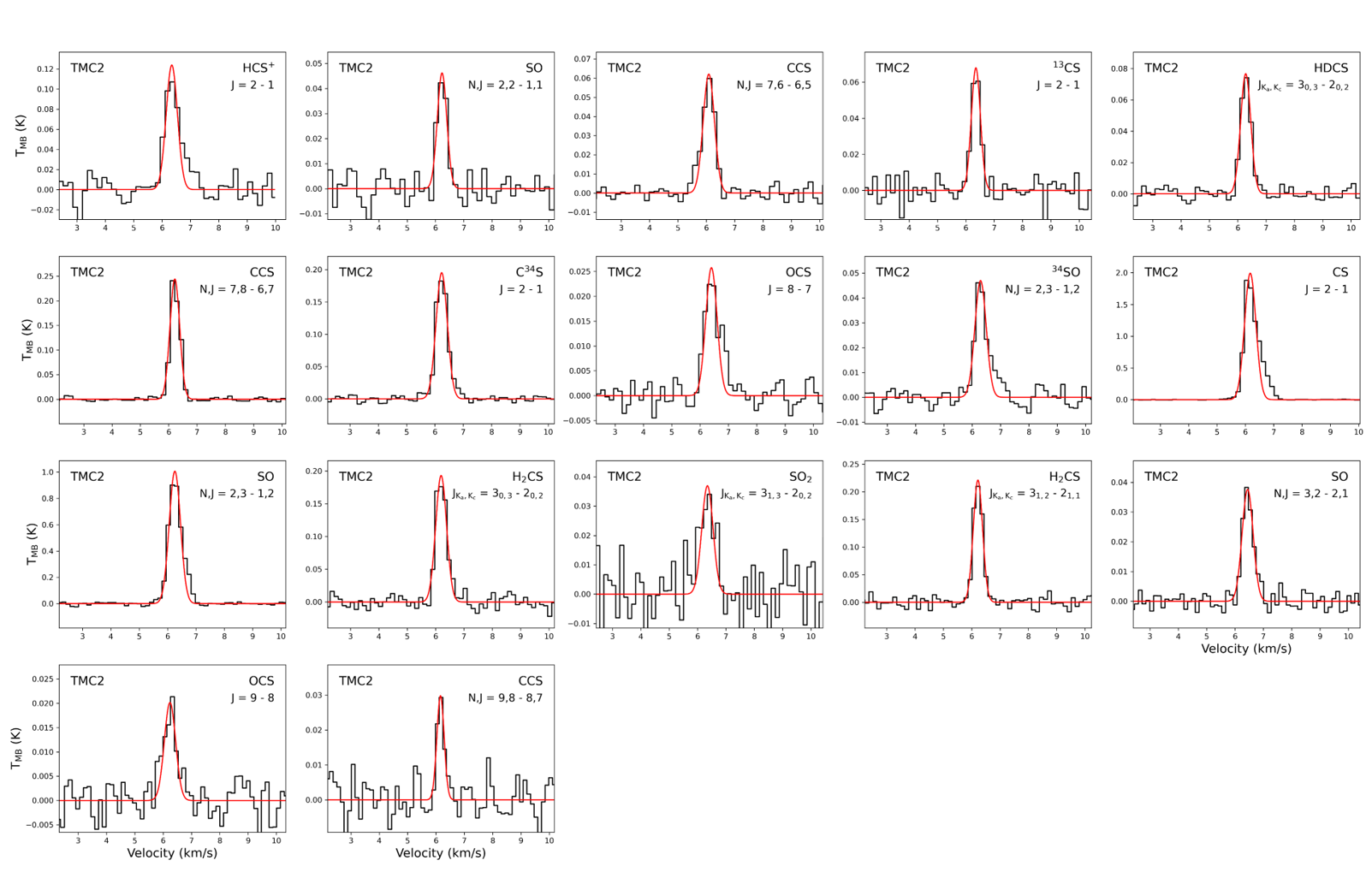}
    \caption{Same as Figure \ref{fig:3_linesL1495AS} but for the pre-stellar core TMC2.}
    \label{fig:A_linesTMC2}
\end{figure*}

\begin{figure*}[h!]
    \centering
    \includegraphics[width=0.95\textwidth]{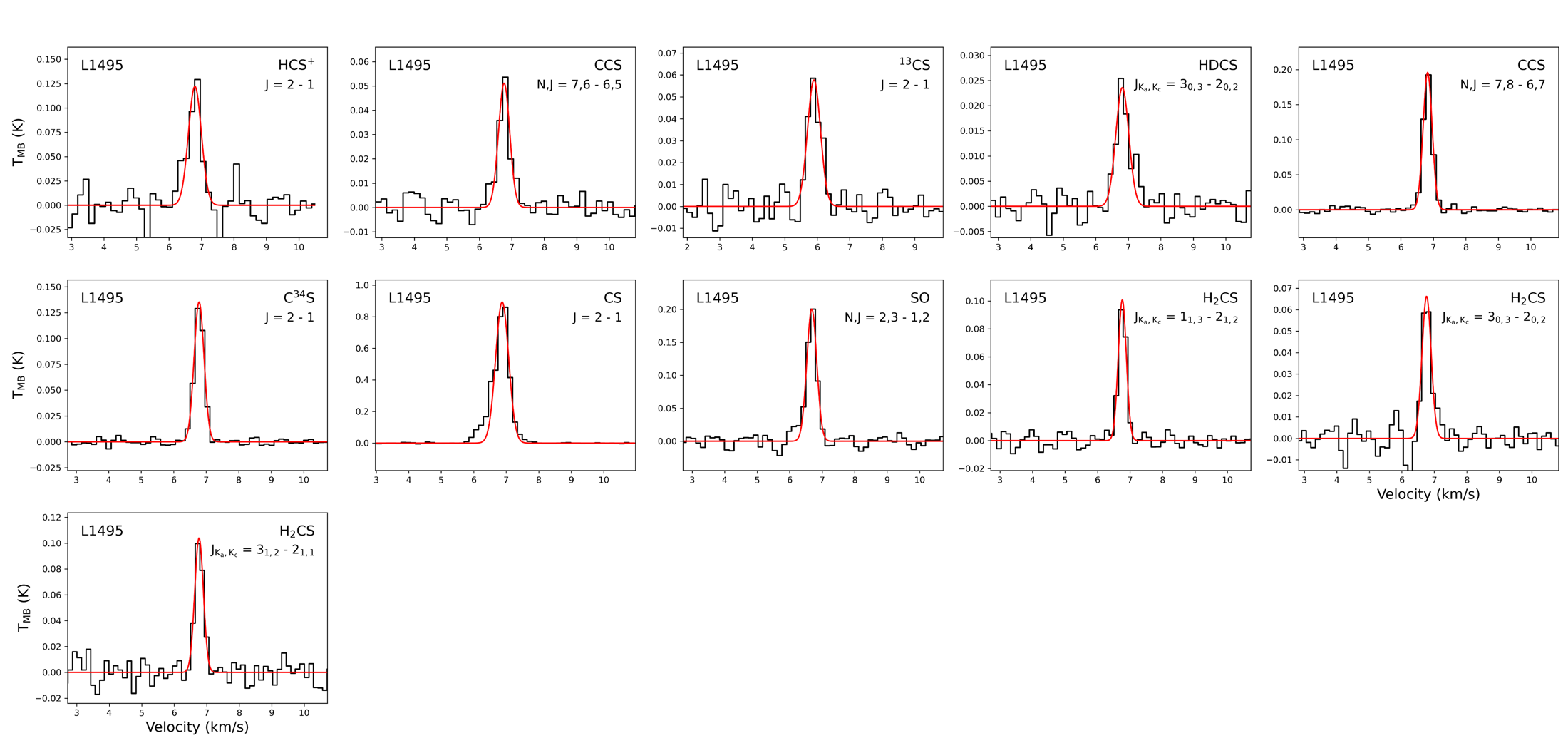}
    \caption{Same as Figure \ref{fig:3_linesL1495AS} but for the starless core L1495.}
    \label{fig:A_linesL1495}
\end{figure*}

\begin{figure*}[h!]
    \centering
    \includegraphics[width=0.95\textwidth]{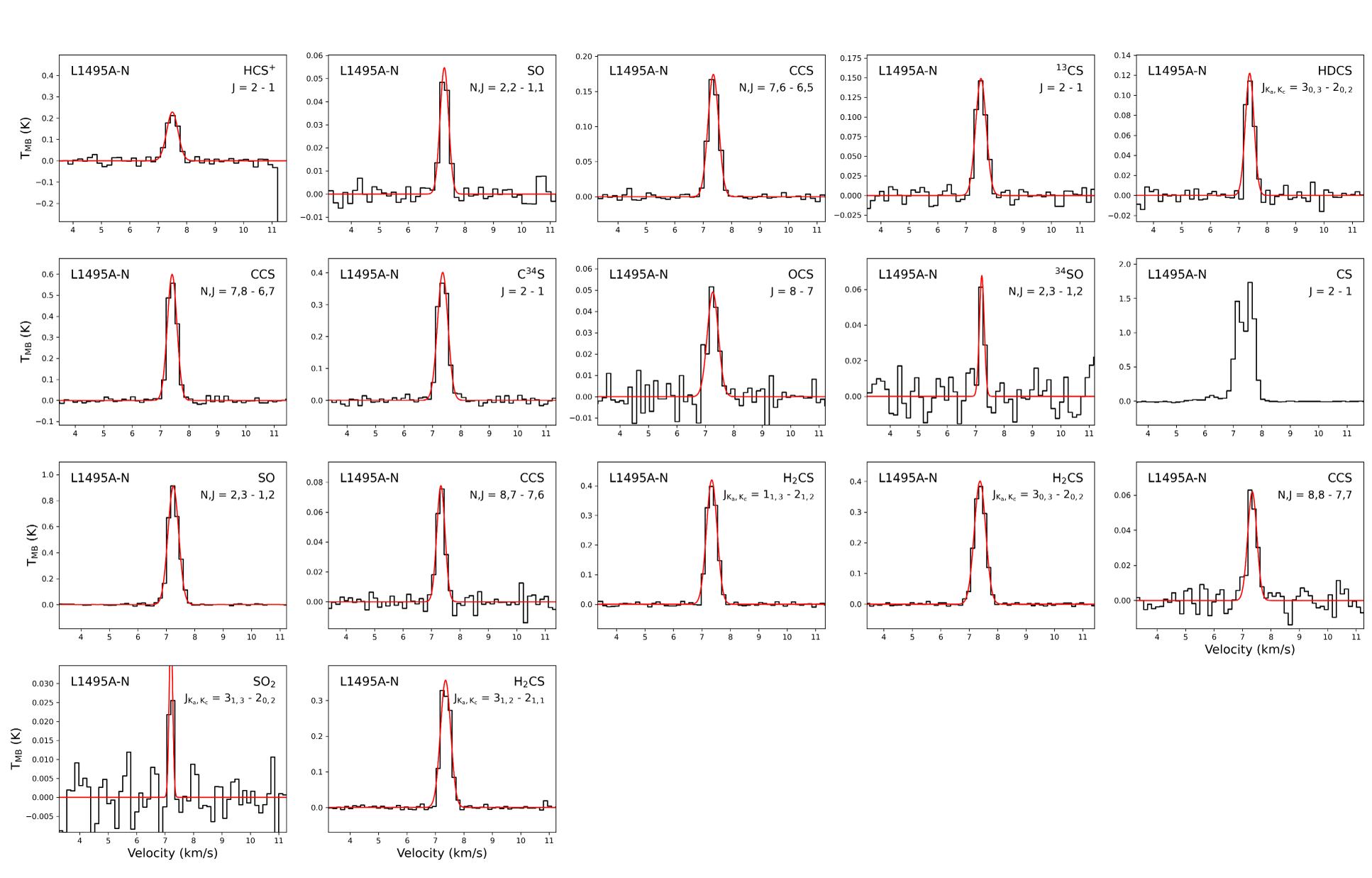}
    \caption{Same as Figure \ref{fig:3_linesL1495AS} but for the starless core L1495A-N. CS shows signs of self-absorption; hence, no Gaussian fit was performed.}
    \label{fig:A_linesL1495AN}
\end{figure*}

\begin{figure*}[h!]
    \centering
    \includegraphics[width=0.95\textwidth]{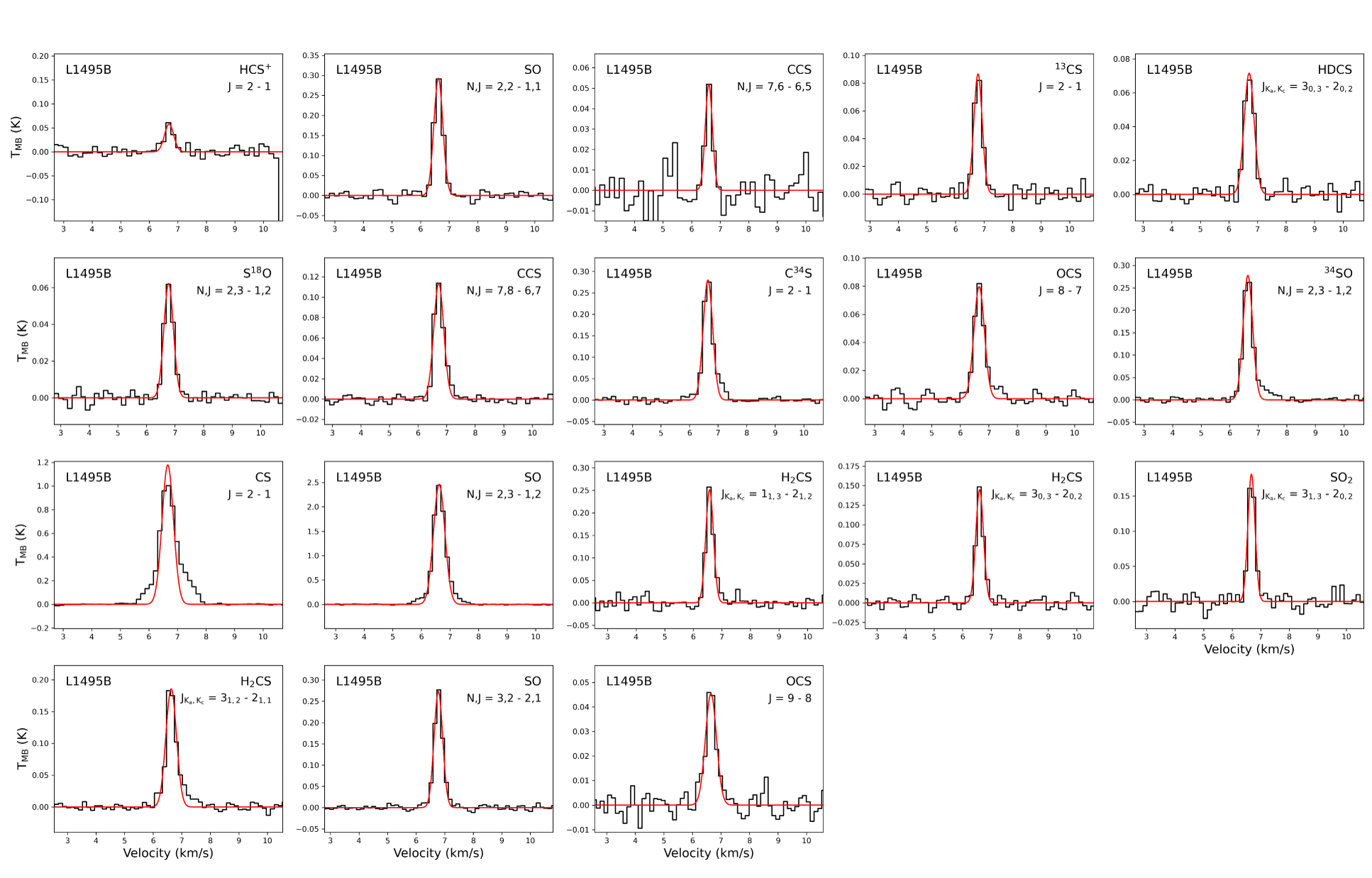}
    \caption{Same as Figure \ref{fig:3_linesL1495AS} but for starless core L1495B. }
    \label{fig:A_linesL1495B}
\end{figure*}

\begin{figure*}[h!]
    \centering
    \includegraphics[width=0.95\textwidth]{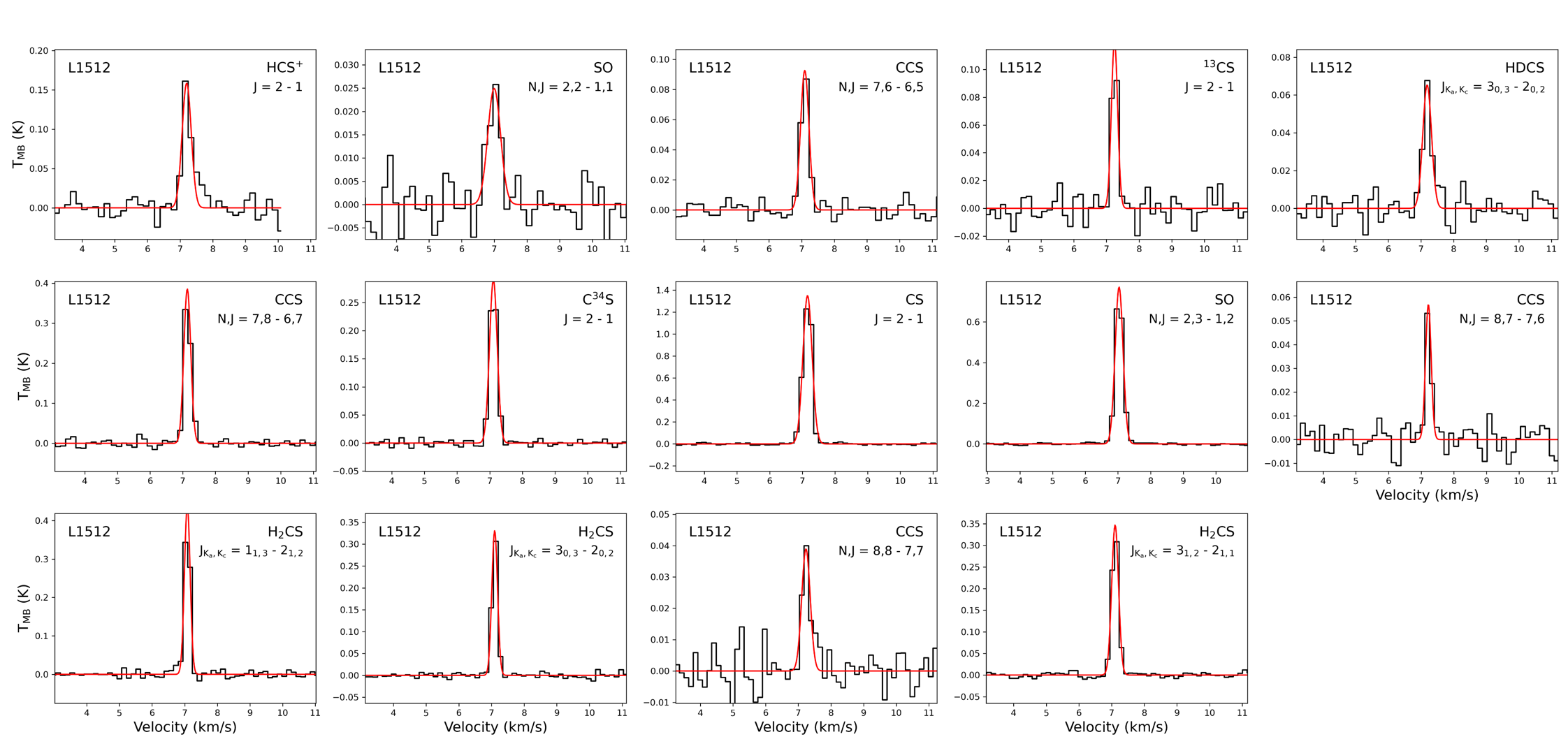}
    \caption{Same as Figure \ref{fig:3_linesL1495AS} but for starless core L1512. }
    \label{fig:A_linesL1512}
\end{figure*}

\begin{figure*}[h!]
    \centering
    \includegraphics[width=0.95\textwidth]{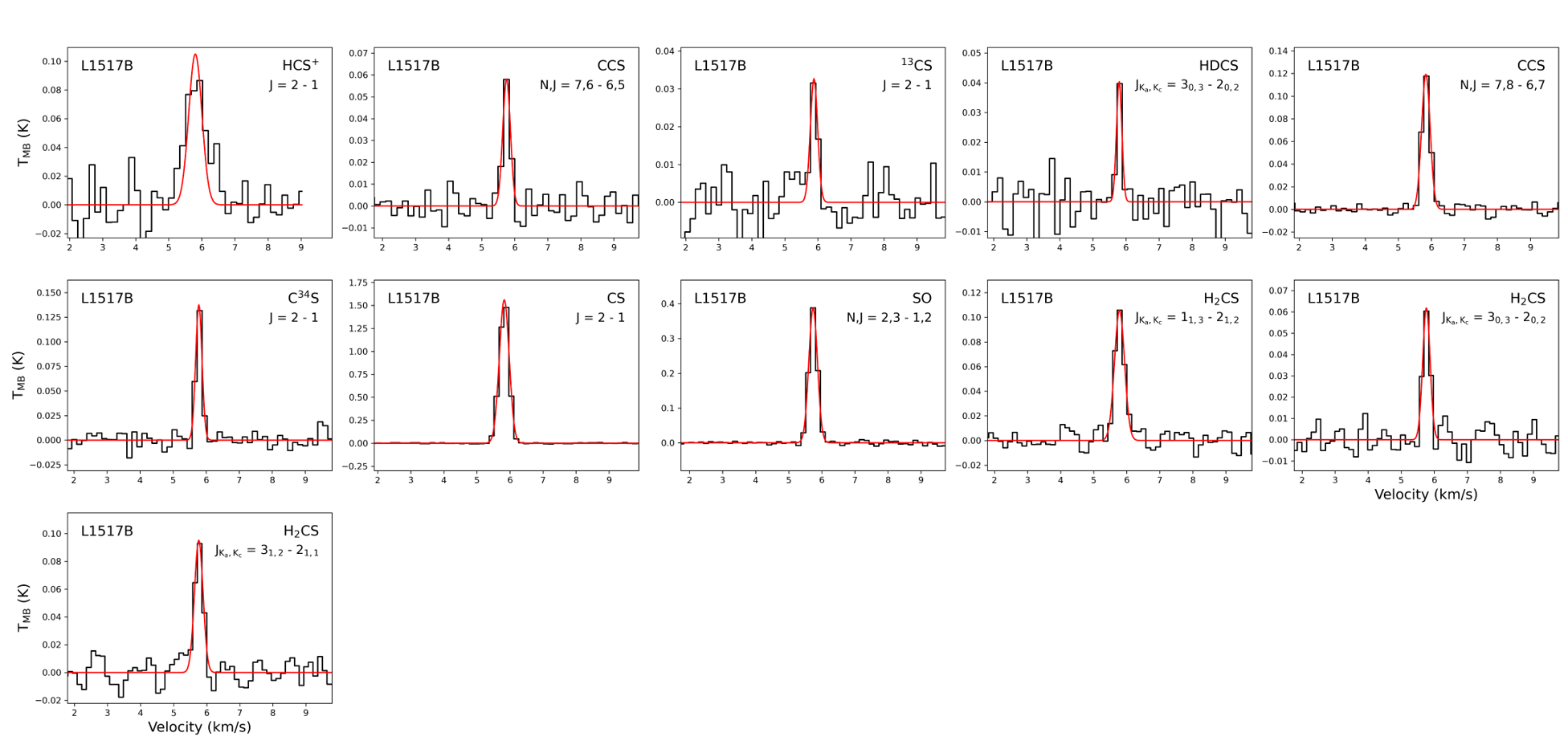}
    \caption{Same as Figure \ref{fig:3_linesL1495AS} but for starless core L1517B. }
    \label{fig:A_linesL1517B}
\end{figure*}

\begin{figure*}[h!]
    \centering
    \includegraphics[width=0.95\textwidth]{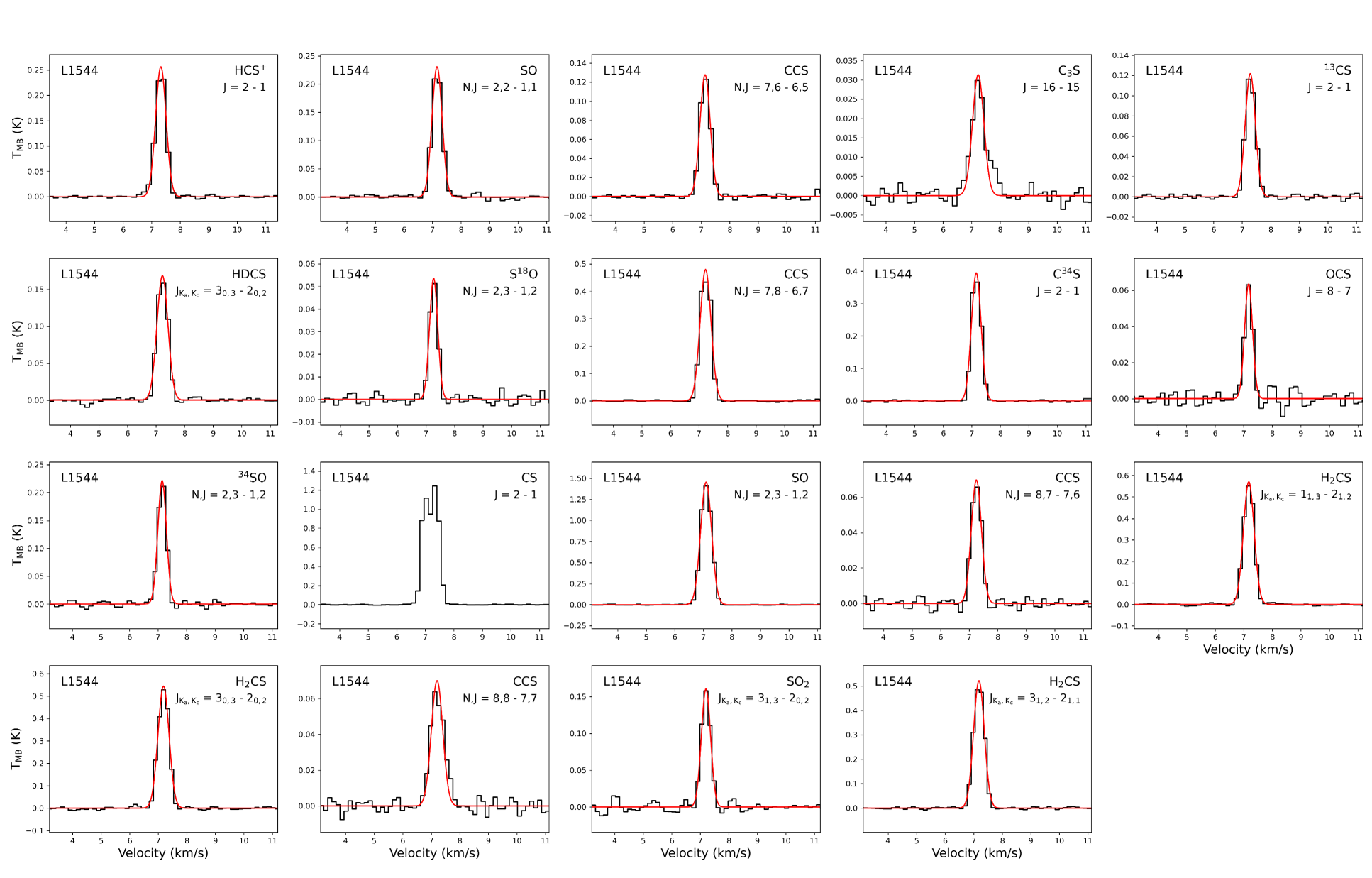}
    \caption{Same as Figure \ref{fig:3_linesL1495AS} but for pre-stellar core L1544. CS shows signs of self-absorption; hence, no Gaussian fit was performed.}
    \label{fig:A_linesL1544}
\end{figure*}

Figures \ref{fig:A_linesCB23} - \ref{fig:A_linesL1544} show the observed spectra of all sulfur-bearing molecules including a Gaussian Fit toward all sources in the sample.

\newpage

\newpage 

\newpage

\section{Line list of the core sample}
\label{sec:AppendixB}
Table B.1, showing the fitting parameters, the optical depth, and column density for the remaining cores, is available in electronic form at the CDS.

\newpage
\section{Column densities with hyperfine structure}
\label{sec:AppendixD}

Since the $^{17}$O isotope has a nonzero nuclear spin, the C$^{17}$O (1-0) transition is split into three different hyperfine components. The magnetic and electric quadrupole moments of the nucleus interact with the rotational motion of the molecule, which causes the hyperfine splitting. The \texttt{CLASS} software contains a method for analyzing these hyperfine components, providing information about their positions and relative intensities. The fit provides the optical depth as the sum of the individual components, with a resulting value of $\tau = 0.1$ indicating an optically thin line. Any line with a $\tau$ above $0.1$ cannot automatically be assumed optically thin, which is why the line has to be corrected with the optical depth correction factor (see Equation \ref{eq:3_correctionfactor}).

Six out of the nine cores show a C$^{17}$O ($J$ = 1-0) transition that can be considered optically thin. For those, we used a three-component Gaussian fit to retrieve the line information. To calculate the column density, we used the sum of the individual areas from the three components as integrated intensity and the line parameters of the un-split C$^{17}$O ($J$ = 1-0) transition together with Equation \ref{eq:3_coldens} under the optically thin assumption. For the excitation temperature, we used \SI{10}{K}, as in the previous analysis. The uncertainties were determined by propagating the integrated-intensity uncertainty.

Table \ref{tab:C_C17Ofits} shows the resulting line parameters for the individual components and the total column density for the transition. As all three components stem from the same molecule, each transition should have a similar linewidth. This was the case for all cores except L1495A-N and L1495B, for which the linewidth had to be fixed to the core's average to ensure consistency. Additionally, L1495B shows a peculiar line feature that resembles a second velocity component. As none of the sulfur-bearing molecules in that core exhibit such a feature, we attempted to fit the three main components as accurately as possible. The resulting value should thus be seen as an approximation.

\renewcommand{\arraystretch}{1.3}
\begin{table*}[h!]
    \centering
    \caption{Line properties of the C$^{17}$O hyperfine triplet, obtained from simultaneous Gaussian fits to the three hyperfine components in the cores TMC2, L1495, L1495A-N, L1495B, L1512, and L1517B.}
    \vspace{-3mm}
    \begin{tabularx}{\textwidth}{lXcccccc}
        \hline\hline
        Source / & Frequency & $T_\text{MB}$ & rms &  $W$  &  v$_\text{LSR}$ &$\Delta$v &  $N_\text{tot}$ \\ 
        Transition      &  (GHz)    & (K)           & (mK)&(K km s$^{-1}$)& (km s$^{-1}$) &(km s$^{-1}$) & ($10^{-14}$ cm$^{-2}$) \\
        \hline        
        \large\textbf{\textit{TMC2}} &  &  &  &  &  & & \\
        $J, F$ = 1,7/2 - 0, 5/2   & 112.36001 &  0.29     &  8   & 0.146 $\pm$    0.003 & 6.354 $\pm$     0.005 & 0.480 $\pm$     0.011 & 5.10  $\pm$ 0.08    \\
        $J, F$ = 1,5/2 - 0, 5/2   & 112.35899 &  0.41     &  8   & 0.199 $\pm$    0.005 & 9.072 $\pm$     0.005 & 0.452 $\pm$     0.012 &        \\
        $J, F$ = 1,3/2 - 0, 5/2   & 112.35879 &  0.23     &  8   & 0.092 $\pm$    0.004 & 9.603 $\pm$     0.008 & 0.385 $\pm$     0.018 &        \\
        \hline

        \large\textbf{\textit{L1495}} &  &  &  &  &  & & \\
        $J, F$ = 1,7/2 - 0, 5/2   & 112.36001 &  0.17     &  6   & 0.084 $\pm$    0.002 & 6.838 $\pm$     0.006 & 0.471 $\pm$     0.018 & 2.98  $\pm$ 0.07    \\
        $J, F$ = 1,5/2 - 0, 5/2   & 112.35899 &  0.21     &  6   & 0.132 $\pm$    0.004 & 9.562 $\pm$     0.007 & 0.571 $\pm$     0.025 &  \\
        $J, F$ = 1,3/2 - 0, 5/2   & 112.35879 &  0.10     &  6   & 0.039 $\pm$    0.003 & 10.132$\pm$     0.013 & 0.363 $\pm$     0.025 &  \\
        \hline
        \large\textbf{\textit{L1495A-N}} &  &  &  &  &  & & \\
        $J, F$ = 1,7/2 - 0, 5/2   & 112.36001 &  0.36     &  14  & 0.170 $\pm$    0.004 & 7.312 $\pm$     0.007 & 0.450 $\pm$     0.000 & 6.06  $\pm$ 0.10    \\
        $J, F$ = 1,5/2 - 0, 5/2   & 112.35899 &  0.46     &  14  & 0.221 $\pm$    0.005 & 10.003$\pm$     0.007 & 0.450 $\pm$     0.000 &        \\
        $J, F$ = 1,3/2 - 0, 5/2   & 112.35879 &  0.27     &  14  & 0.127 $\pm$    0.005 & 10.494$\pm$     0.014 & 0.450 $\pm$     0.000 &        \\
        \hline
        \large\textbf{\textit{L1495B}} &  &  &  &  &  & & \\
        $J, F$ = 1,7/2 - 0, 5/2   & 112.36001 &  0.51     &  23          & 0.217 $\pm$    0.007 & 7.036 $\pm$     0.009 & 0.400 $\pm$     0.000 &  8.57 $\pm$ 1.35        \\
        $J, F$ = 1,5/2 - 0, 5/2   & 112.35899 &  0.82     &  23          & 0.351 $\pm$    0.007 & 9.768 $\pm$     0.006 & 0.400 $\pm$     0.000 &         \\
        $J, F$ = 1,3/2 - 0, 5/2   & 112.35879 &  0.39     &  23          & 0.167 $\pm$    0.007 & 10.300$\pm$     0.012 & 0.400 $\pm$     0.000 &         \\
        \hline
        \large\textbf{\textit{L1512}} &  &  &  &  &  & & \\
        $J, F$ = 1,7/2 - 0, 5/2   & 112.36001 &  0.38     &  16  & 0.095 $\pm$    0.005 & 7.122 $\pm$     0.005 & 0.232 $\pm$     0.011 &  3.45 $\pm$ 0.08    \\
        $J, F$ = 1,5/2 - 0, 5/2   & 112.35899 &  0.60     &  16  & 0.137 $\pm$    0.004 & 9.831 $\pm$     0.003 & 0.216 $\pm$     0.007 &        \\
        $J, F$ = 1,3/2 - 0, 5/2   & 112.35879 &  0.29     &  16  & 0.064 $\pm$    0.004 & 10.367$\pm$     0.007 & 0.210 $\pm$     0.013 &        \\
        \hline
        \large\textbf{\textit{L1517B}} &  &  &  &  &  & & \\
        $J, F$ = 1,7/2 - 0, 5/2   & 112.36001 &  0.19     &  15  & 0.058 $\pm$    0.004 & 5.788 $\pm$     0.011 & 0.288 $\pm$     0.023 &  2.17 $\pm$ 0.09    \\
        $J, F$ = 1,5/2 - 0, 5/2   & 112.35899 &  0.26     &  15  & 0.086 $\pm$    0.005 & 8.491 $\pm$     0.008 & 0.305 $\pm$     0.018 &        \\
        $J, F$ = 1,3/2 - 0, 5/2   & 112.35879 &  0.13     &  15  & 0.041 $\pm$    0.005 & 9.028 $\pm$     0.016 & 0.304 $\pm$     0.041 & \\
        \hline

    \end{tabularx}
    
\tablefoot{The line properties were derived using a Gaussian fit with \texttt{CLASS}. The last column represents the total column density of C$^{17}$O.}

    \label{tab:C_C17Ofits}
\end{table*}
\renewcommand{\arraystretch}{1.0}

\begin{figure*}[htbp]
    \centering

    \begin{subfigure}{0.33\textwidth}
        \centering
        \includegraphics[width=\linewidth]{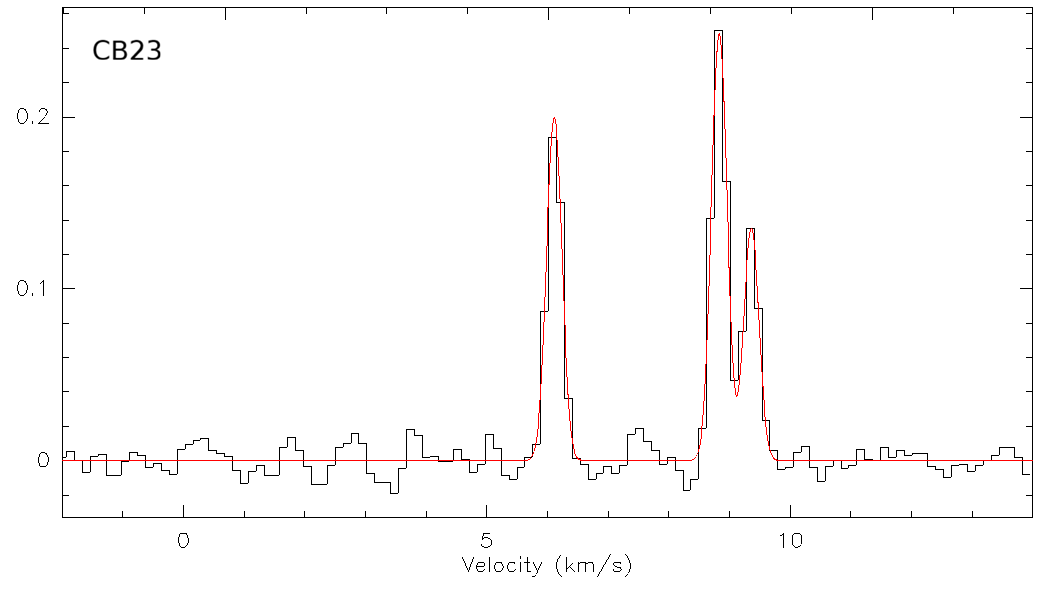}
        \label{fig:sub1}
    \end{subfigure}
    \hfill
    \begin{subfigure}{0.33\textwidth}
        \centering
        \includegraphics[width=\linewidth]{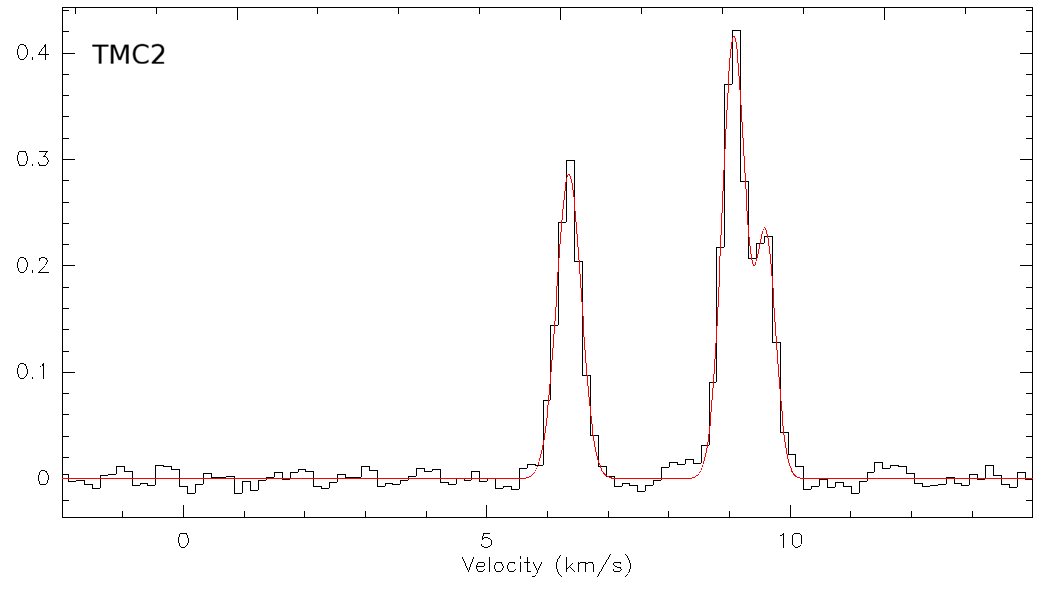}
        \label{fig:sub2}
    \end{subfigure}
    \hfill
    \begin{subfigure}{0.33\textwidth}
        \centering
        \includegraphics[width=\linewidth]{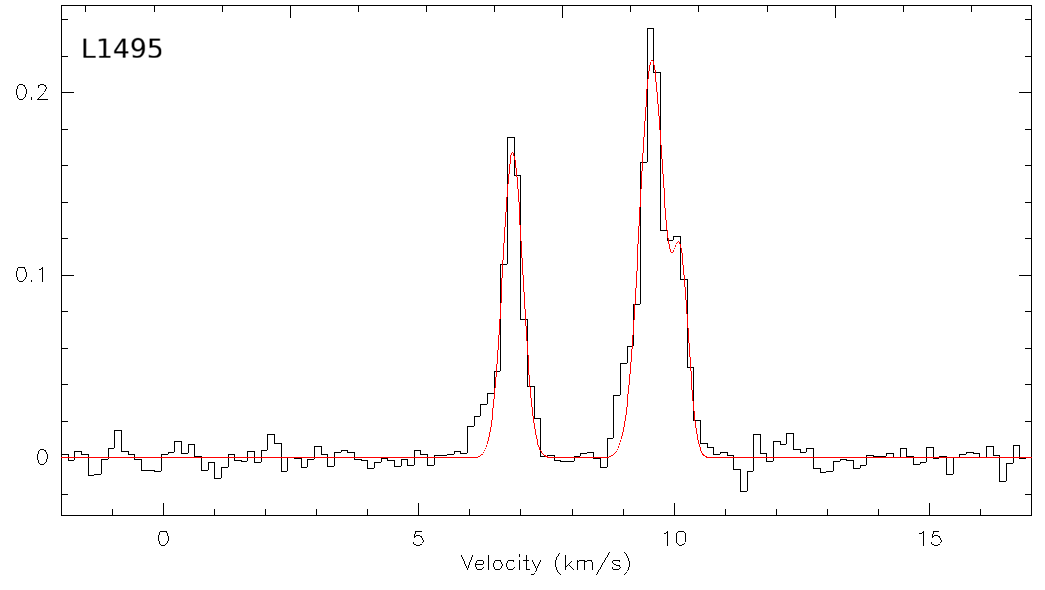}
        \label{fig:sub3}
    \end{subfigure}

    \vskip\baselineskip

    \begin{subfigure}{0.33\textwidth}
        \centering
        \includegraphics[width=\linewidth]{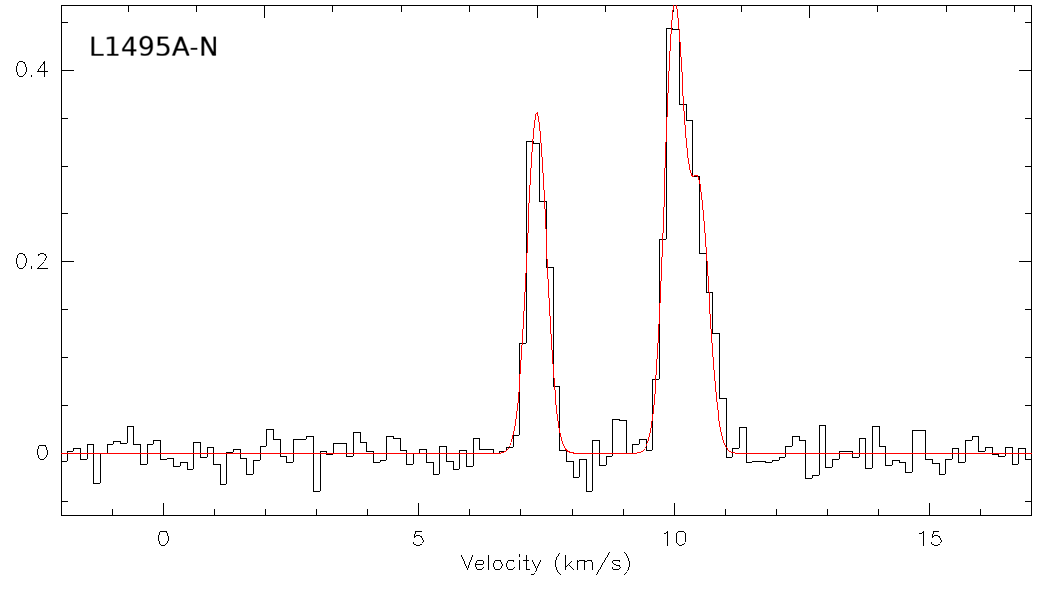}
        \label{fig:sub4}
    \end{subfigure}
    \hfill
    \begin{subfigure}{0.33\textwidth}
        \centering
        \includegraphics[width=\linewidth]{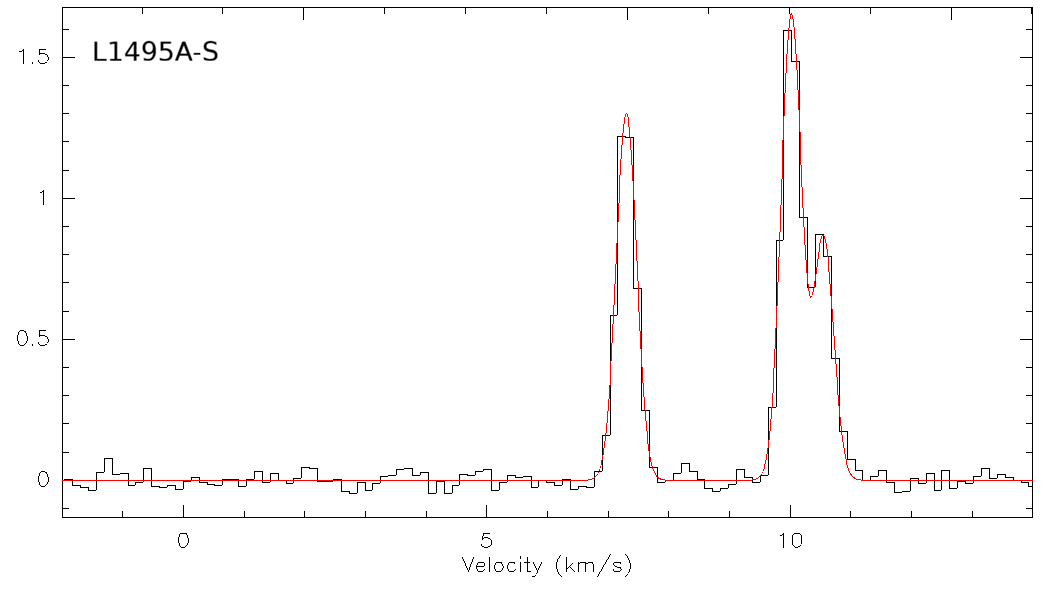}
        \label{fig:sub5}
    \end{subfigure}
    \hfill
    \begin{subfigure}{0.33\textwidth}
        \centering
        \includegraphics[width=\linewidth]{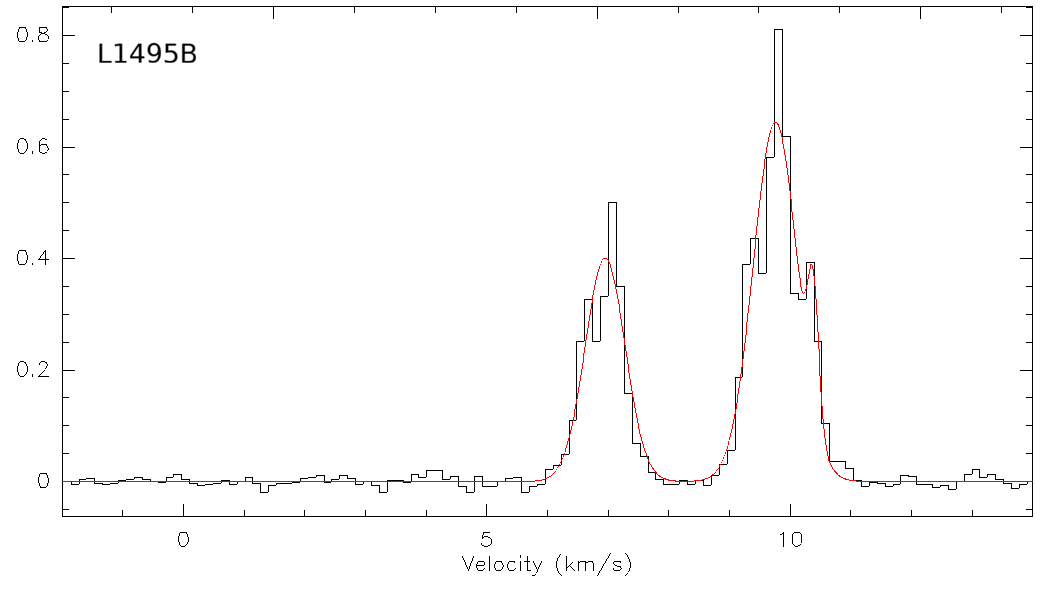}
        \label{fig:sub6}
    \end{subfigure}

    \vskip\baselineskip

    \begin{subfigure}{0.33\textwidth}
        \centering
        \includegraphics[width=\linewidth]{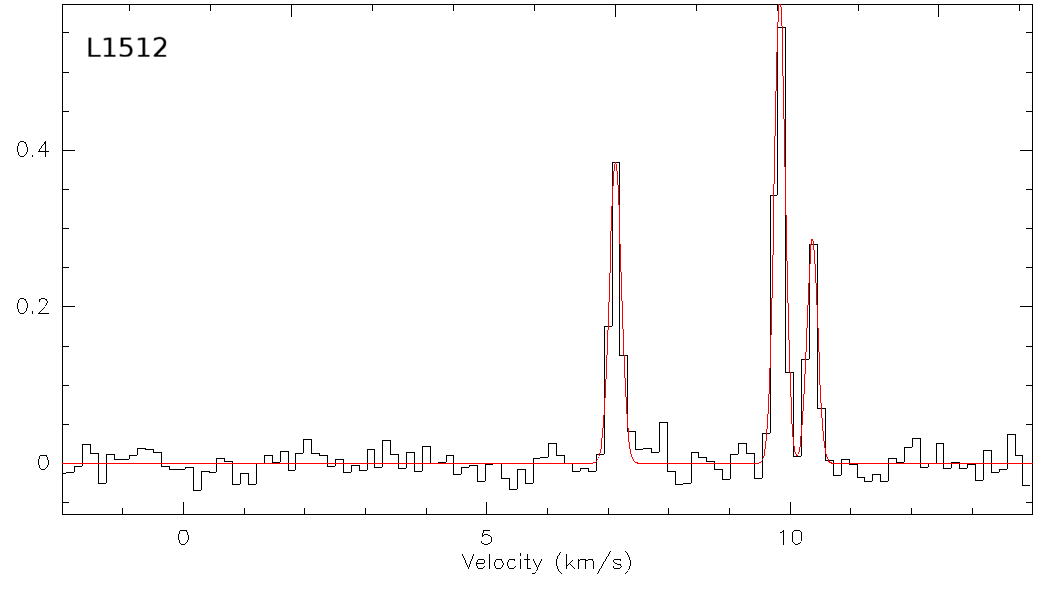}
        \label{fig:sub7}
    \end{subfigure}
    \hfill
    \begin{subfigure}{0.33\textwidth}
        \centering
        \includegraphics[width=\linewidth]{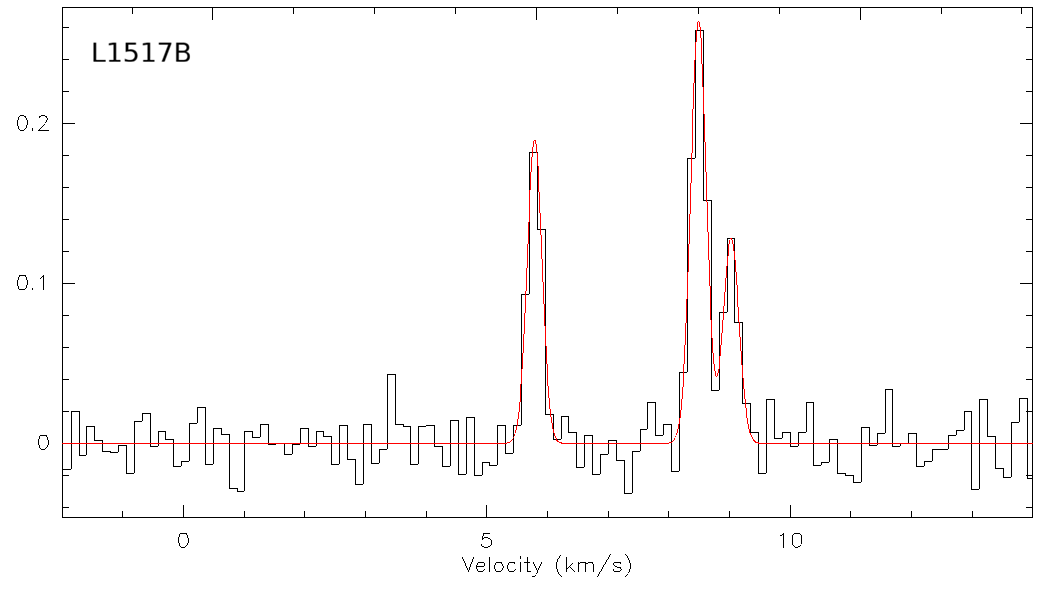}
        \label{fig:sub8}
    \end{subfigure}
    \hfill
    \begin{subfigure}{0.33\textwidth}
        \centering
        \includegraphics[width=\linewidth]{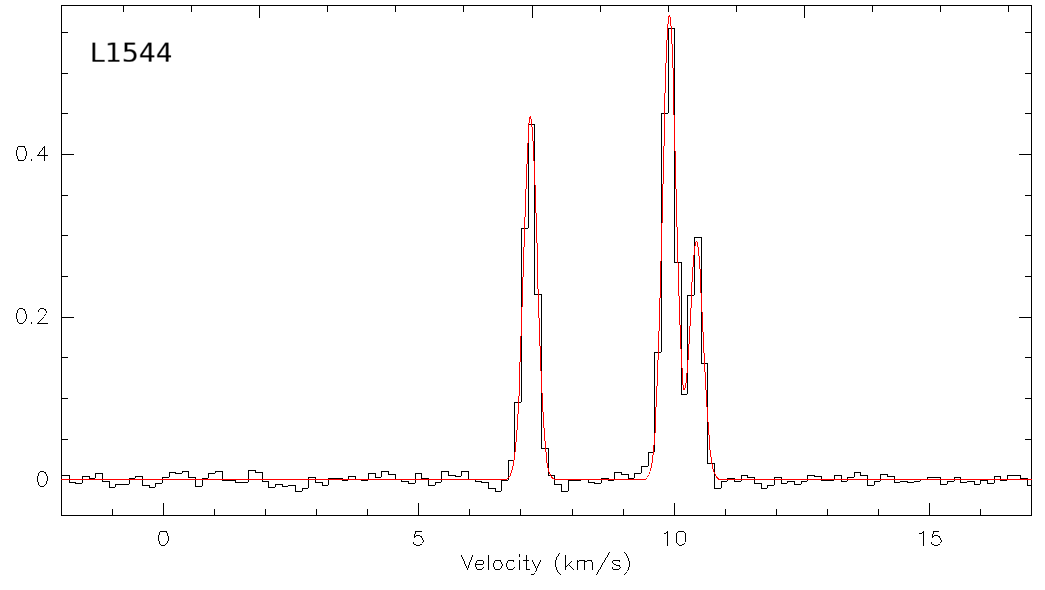}
        \label{fig:sub9}
    \end{subfigure}

    \caption{Spectra of the C$^{17}$O line observed toward the different sources in black. The red line indicates the best fit line for the specified source in the upper-left corner of each panel.}
    \label{fig:C_FitsCO}
\end{figure*}

For CB23, L1495A-S, and L1544, the fitted optical depths exceed 0.1; therefore, we needed to correct them. When exhibiting non-optically thin hyperfine structure, we cannot simply assume a constant excitation temperature, as was done in the previous analysis. Instead, the hyperfine structure (\texttt{hfs}) fit allows us to calculate $T_\text{ex}$ from the fit parameters themselves via
\begin{equation}
    T_\text{ex} = \frac{h \nu / k_\text{B}}{ln \left( \frac{h \nu / k_\text{B}}{(T_\text{MB}\cdot \tau)/(\tau_\text{main}f) + J (T_\text{bg})} + 1 \right) },
\end{equation}
where $T_\text{MB}\cdot \tau$ is the product of the intensity and the optical depth summed over all three components. In the \texttt{hfs} fitting, this corresponds to the first output parameter, while $\tau_\text{MB}$ corresponds to the last. The calculated excitation temperature can then be used to determine the total column density via
\begin{equation}
    N_\text{tot} = \tau_\text{main} \sqrt{\frac{16 \pi^3}{\ln{(2)}}} \frac{\nu^3 Q(T_\text{ex}) \Delta \text{v}}{c^3 A_\text{u} g_\text{u}}  \frac{e^{\frac{E_\text{u}}{k_\text{B}T_\text{ex}}} }{e^{\frac{h \nu}{k_\text{B}T_\text{ex}}}-1}.
\end{equation}
Table \ref{tab:C_hfs_results} shows the results of the \texttt{hfs} fitting and the column densities for C$^{17}$O in CB23, L1495A-S, and L1544. The uncertainties were determined using the Gaussian error propagation method. The resulting fits are presented in Figure \ref{fig:C_FitsCO}.

\renewcommand{\arraystretch}{1.3}
\begin{table*}[]
    \centering    
    \caption{\texttt{hfs} fitting results and column densities of the C$^{17}$O hyperfine triplet in CB23, L1495A-S, and L1544. }
    \vspace{-3mm}
    \begin{tabularx}{\textwidth}{@{}lYYYYY@{}}
    \hline\hline
    Core   & $T_{\text{MB}} \cdot \tau$ & v$_{\text{LSR}}$ & $\Delta$v & $\tau$ & $N_\text{tot}$ \\
            &  (K)                       &  (km/s)           &  (km/s)    &         & ($10^{-14}$ cm$^{-2}$) \\
    \hline
    CB23    &  0.669 $\pm$ 0.067 & 6.11 $\pm$ 0.003 & 0.287 $\pm$ 0.010 & 0.789 $\pm$ 0.578 & 3.76 $\pm$ 2.83\\
    L1495A-S&  4.07 $\pm$ 0.173 & 7.30 $\pm$ 0.002 & 0.388 $\pm$ 0.006 & 0.378 $\pm$ 0.231 & 18.17 $\pm$ 15.40 \\
    L1544   &  1.53 $\pm$ 0.083 & 18.7 $\pm$ 0.002 & 0.293 $\pm$ 0.006 & 0.782 $\pm$ 0.314 & 5.50 $\pm$ 2.40\\
    \hline
    \end{tabularx}
    \label{tab:C_hfs_results}
\end{table*}

\end{appendix}

\end{document}